\documentclass[useAMS,usenatbib,pslatex]{mn2e}

\title[Monthly Notices: \LaTeXe\ guide for authors]
  {A comprehensive photometric study of circumnuclear star forming rings I: the sample.}
\author[Mar \'{A}lvarez \'{A}lvarez et al.]
  {Mar \'{A}lvarez - \'{A}lvarez,$^{1,2}$ \thanks{e-mail:malvar1@ing.uc3m.es}, Angeles I. D\'{i}az,$^{2}$, Elena Terlevich$^{3}$\thanks{Visiting astronomer at IoA, University of
  Cambridge, UK}  and 
Roberto Terlevich$^{3,4}$\\
 $^1$ Departamento de F\'{\i}sica, Universidad Carlos III, Getafe, Spain.\\
  $^2$ Departamento de F\'{\i}sica Te\'{o}rica, Universidad Aut\'{o}noma de Madrid, Cantoblanco, 28049-Madrid, Spain.\\
  $^3$ Instituto Nacional de Astrof\'\i sica \'Optica y Electr\'onica, L.E.~Erro No. 1, Santa Mar\'\i a Tonantzintla,  Puebla, M\'exico\\
  $^4$ Institute of Astronomy,University of Cambridge, Madingley Road, Cambidge CB3 0HA, UK\\}

\date{Released 2012 Xxxxx XX}

\pagerange{\pageref{firstpage}--\pageref{lastpage}} \pubyear{2011}

\def\LaTeX{L\kern-.36em\raise.3ex\hbox{a}\kern-.15em
    T\kern-.1667em\lower.7ex\hbox{E}\kern-.125emX}

\usepackage{graphicx}

\begin{document}

\label{firstpage}

\maketitle{}

\begin{abstract}
We present photometry in U, B, V, R and I continuum bands and in H$\alpha$ and H$\beta$ emission lines for a sample of 336 circumnuclear star forming regions (CNSFR) located in early type spiral galaxies with different levels of activity in their nuclei. They are nearby galaxies, with distances less than 100 Mpc, 60\% of which are considered as interacting objects. 
	This survey of 20 nuclear rings aims to provide insight into their star formation properties as age, stellar population and star formation rate. Extinction corrected H$\alpha$ luminosities range from $1.3\times  10^{38}$ to $4\times 10^{41} erg s^{-1}$, with most of the regions showing values between 39.5 $\leq log L(H\alpha) \leq$ 40, which implies masses for the ionizing clusters higher than $2\times  10^{5} M_\odot $.
	H$\alpha$ and H$\beta$ images have allowed us to obtain an accurate measure of extinction. We have found an average value of A$_V$  = 1.85 magnitudes. (U-B) colour follows a two maximum distribution around (U-B)$ \simeq$ -0.7, and -0.3; (R-I) also presents a bimodal behaviour, with maximum values of 0.6 and 0.9. Reddest (U-B) and (R-I) regions appear in non-interacting galaxies. Reddest (R-I) regions lie in strongly barred galaxies.
	For a significant number of HII regions the observed colours and equivalent widths are not well reproduced by single burst evolutionary theoretical models. 
	
\end{abstract}


\section{Introduction}
	 
	 The existence of circumnuclear rings, defined as circular or near circular structures of enhanced star formation around the nucleus of some galaxies, was first noted by S\'{e}rsic and Pastoriza in 1967,
when they realized that the central parts of some peculiar galaxies showed a rate of star 
formation much higher than it was expected in objects of their characteristics	(\citealt{thy:76}).
Typically located within the central kiloparsecs of galaxies and surrounding their nucleus or a central region in which luminous matter is not detected, the study of these rings, as places where star formation seems to be increased (although some dusty red nuclear rings have been reported in elliptical and SO galaxies (\citealt{com:10}; \citealt{lau:05})), has been proved to be of great importance to understand the internal dynamics of galaxies and their secular evolution (\citealt{maz:08}).

%
%

\begin{figure*}

\includegraphics[scale=0.9]{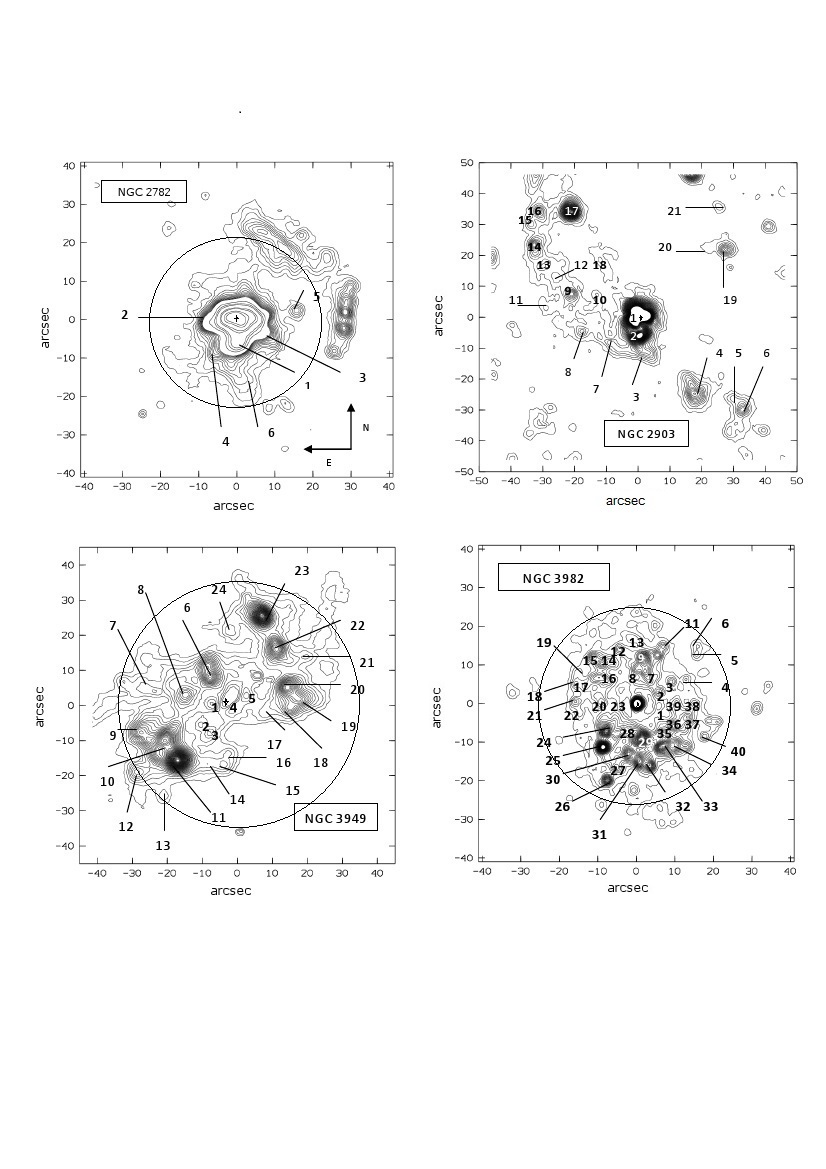} 

\end{figure*}

\begin{figure*}

\includegraphics[scale=0.9]{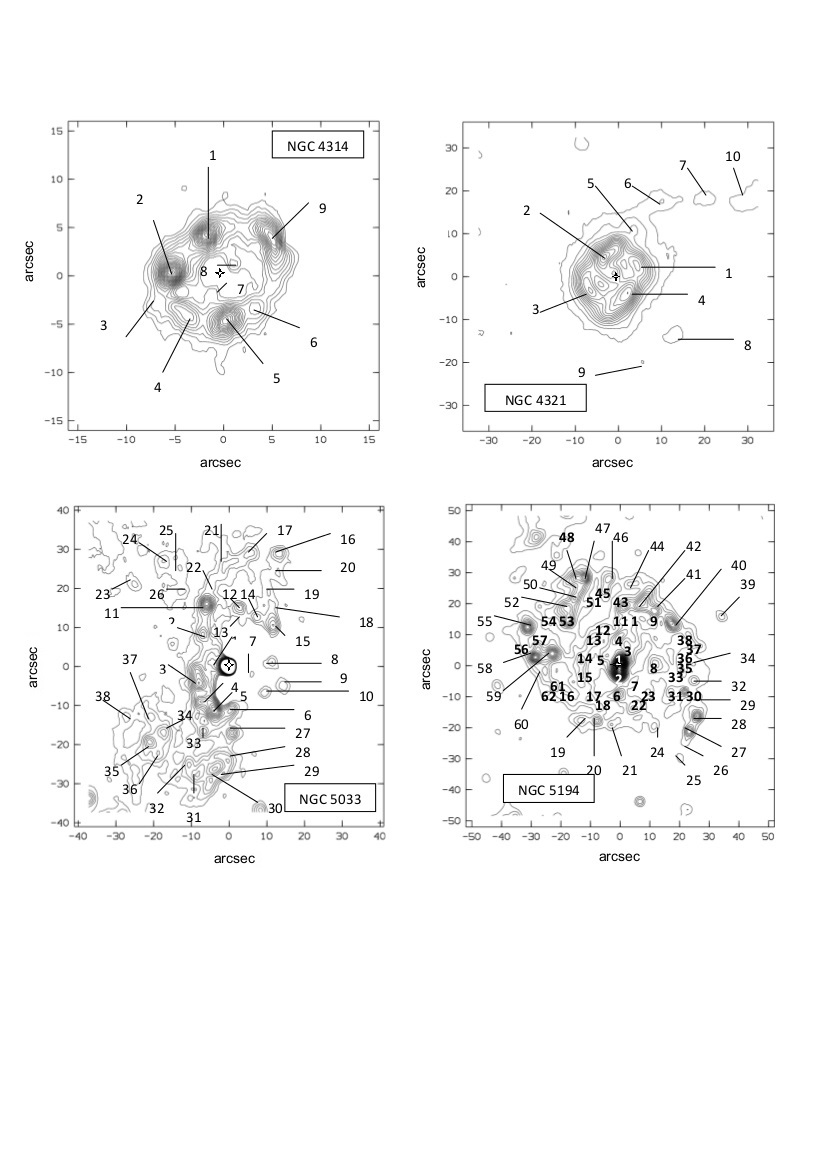}

\end{figure*}

\begin{figure*}

\includegraphics[scale=0.9]{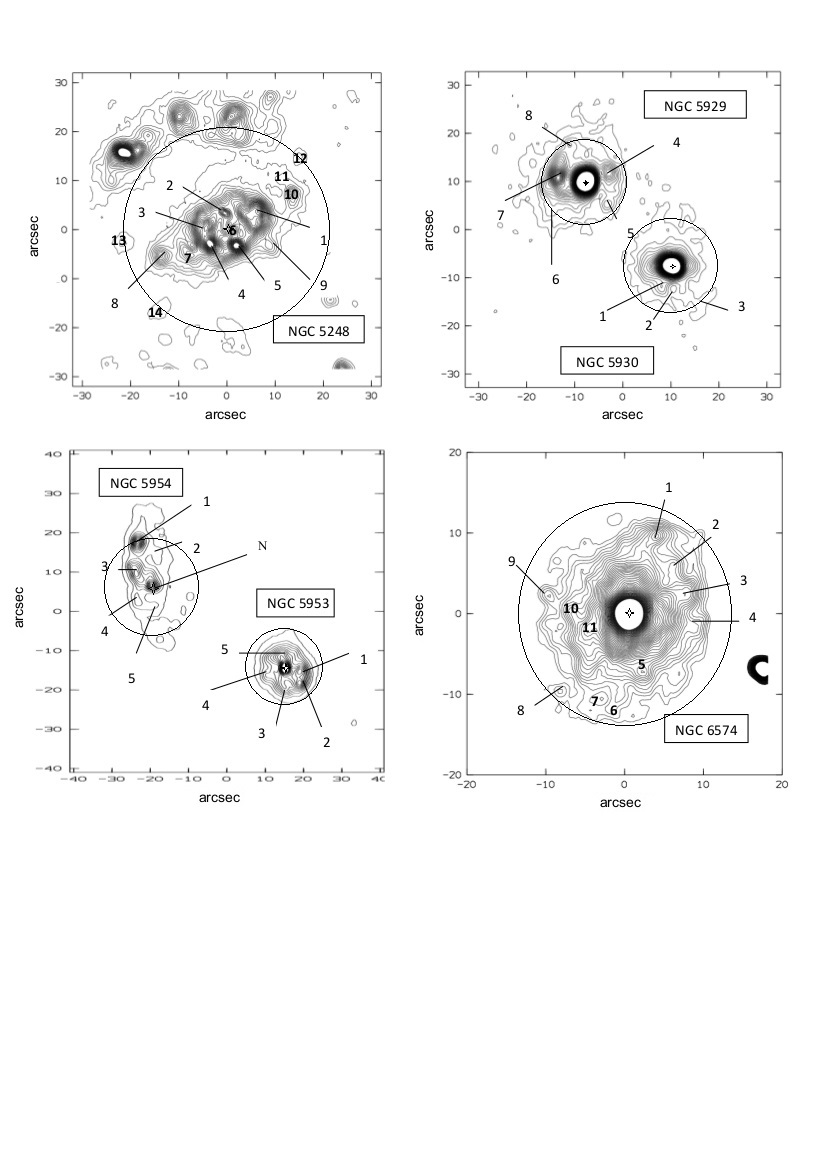}

\end{figure*}

\begin{figure*}

\includegraphics[scale=0.6]{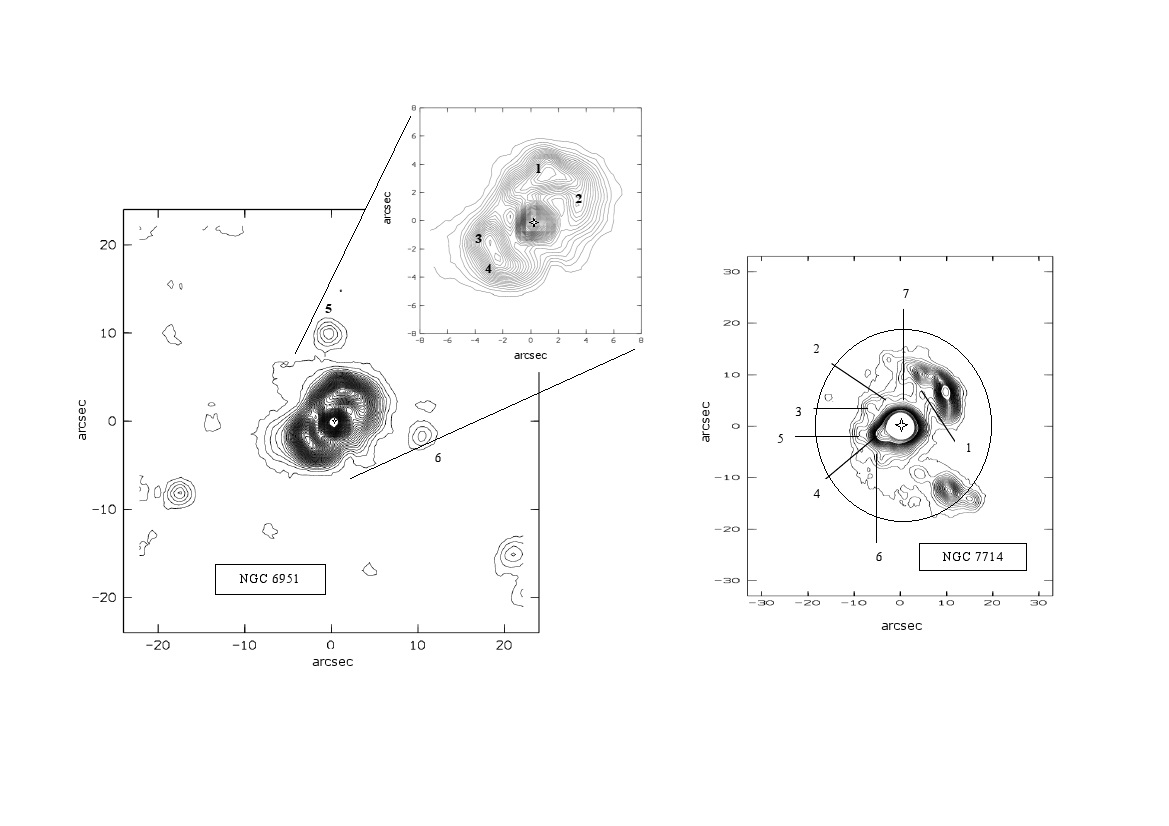} 
\caption{Isophotal maps  $H\protect\alpha +[NII]$ (see text) of the galaxy sample with HII region identification labels. North is to the top and East to the left in all frames. Drawn contour levels are not homogeneous among galaxies nor related to galaxy physical parameters. They have been chosen for a better identification of individual regions only. In some galaxies, limiting radius, as computed in section 4 is included. This radius is not shown when its size is similar to the axis box. The nucleus is also marked with a cross.} 

\label{contours}
\end{figure*}


Due to this reason, many studies can be found in the literature concerning ring galaxies from different points of view: chemical, dynamical or relative to stellar populations (\citealt{elm:94}; \citealt{elm:97}; \citealt{ken:89}; \citealt{hof:97}; \citealt{alb:01}; \citealt{com:10}).  
It has been shown that nuclear rings contain a mixture of neutral and ionized gas and dust with average masses of $10^{9}$ M$_\odot$ (\citealt{rub:97}). 
Also the common association of star forming regions and resonances suggests a close link between star formation and the dynamical behaviour of galaxies. Depending on the longevity of the ring, its ability to replenish itself with gas and its effectiveness as a barrier for gas inflow to the nucleus, the ring will play an increasingly important role in the secular evolution of the galaxy.\\
The relationship of these structures with bars and non axisymmetric features in the
inner parts of galaxies has been studied by \cite{com:85} and \cite{ath:92} among others. Theoretical models (\citealt{woo:13}; \citealt{mac:04} a) b); \citealt{reg:03}) in which the behaviour of gas in the galactic potential is simulated, show a relationship between the presence of a circumnuclear ring and the existence of resonances at the ends of a nuclear bar,
since in these places the gas loses angular momentum and stops its infall towards the interior
of the potential well of the galaxy. These nuclear rings are the locations
of strong density enhancements of gas inside which massive star formation is often detected.

The general objective of this work is to understand the nature and evolution processes that take place within HII regions located in circumnuclear rings of galaxies and characterize the star population present in each knot: its mass, age, extinction parameters, etc.
We intend to determine whether the HII region properties are linked to galactic parameters such as Hubble type, nuclear activity or environment; finding out the role that circumnuclear activity plays in the wider frame of galactic star formation.

To achieve these aims we present photometry in U, B, V, R and I broad continuum bands and in narrow H$\alpha$ and H$\beta$ filters for a sample of 336 circumnuclear star forming regions (CNSFR) located in 20 early type spiral galaxies with different levels of activity in their nuclei.  The sample is described in \S 2. Information on the observations and data reduction can be found in \S 3. The ring definition adopted  and data analysis are explained in \S 4; \S 5 is devoted to the presentation of results and their discussion. Finally, in \S 6 we present a brief summary and the main conclusions of the work. 

%
%

\begin{table*}
 \begin{minipage}{100mm}
 \caption{Characteristics of the galaxy Sample: nuclear type is from \protect \citealt{ver:06}, if the galaxy is present in the Catalog. Otherwise nuclear type is from the NED (*)  } \label{sample}
 \begin{tabular}{@{}cccccc}
  \bf Galaxy & \bf Morphological & \bf angular scale & \bf Nuclear & & 
  \bf D (Mpc) \\
           &   \bf Type $^{1}$    &\bf(pc/$^{\prime\prime}$)    &  \bf Type   &                         \\
 \hline
NGC 1068 &(R)SA(rs)b      & 88      &Sy1               &    &18.1$^{2}$  \\

  NGC 2782 &SAB(rs)a         &165     &Sy2                &  &34.0  \\

  NGC 2903 &SAB(rs)bc             &42       &HII                &   &8.6$^{3}$  \\

  NGC 3310 &SAB(r)bc pec   &61       & Starburst    &   &12.5 \\

  NGC 3949 &SA(s)bc           &52       &--                  &   &10.8  \\

  NGC 3982 &SAB(r)b           &72       &Sy1.9               &   &14.8  \\

  NGC 4314 &SB(rs)A           &49       &LINER            &   &10.0$^{4}$  \\

  NGC 4321 &SAB(s)bc         &83       &HII                &   &17.1$^{5}$  \\

  NGC 5033 &SA(s)c             &91       &Sy1.8               &   &18.7$^{6}$  \\

  NGC 5194 &SA(s)bc pc      &41       &Sy2           &     &8.4$^{7}$  \\
 
  NGC 5248 &(R)SB(rs)bc     &110     &HII               &    &22.7$^{6}$  \\

  NGC 5929 &Sab: pec         &181     &LINER              &     &37.3  \\

  NGC 5930 &SAB(rs)b pec  &181     &HII               &    &37.3  \\

  NGC 5953 &Saa:pec          &130     &Sy2              &     &26.8$^{8}$  \\

  NGC 5954 &SAB(rs)cd       &130     &Sy2              &     &26.8$^{8}$  \\

  NGC 6574 &SAB(rs)bc       &158     &Sy2              &     &32.6  \\

  NGC 6951 &SAB(rs)bc       &117     &Sy2              &     &24.1$^{6}$  \\

  NGC 7177 & SAB(r)b          & 74     & LINER          &    & 15.3$^{4}$ \\

  NGC 7469 & (R')SAB(rs)a  & 315    & Sy1.5             &     & 65.0 \\

  NGC 7714 &SB(s)b pec     &180     &HII                &     &37.2 \\

 \hline
 \end{tabular}
  Notes to table 1:\\
1.- Morphological type is taken from the 3$^{rd}$ Catalog of Bright Galaxies (RC3; de Vaucouleurs et al. 1991).\\
2.- Telesco, C.M., y Harper, D.A. 1980 ApJ. 235, 392. \\
3.- Bottinelli et al. 1984. \\
4.- Garc\'{i}a Barreto, J.A. 1991 A\&A. 244, 272.\\
5.- Freedman, W. L. \textit{et al}., 1994 ApJ. 435, L31.\\
6.- Tully, R.B. 1988, Nearby Galaxies Catalog, (Cambridge Univ. Press: Cambridge).\\
7.- Feldmeier, J.J., Ciardullo, R., Jacoby, G.H., 1997 ApJ., 479, 231.\\
8.- Gonz\'{a}lez-Delgado, R.M. \& Perez, E., 1996 MNRAS. 281, 781.\\
(*)The National Extragalactic Database (NED), is operated by the Jet Propulsion Laboratory, California Institute of Technology, under contract with the National Aeronautics and Space Administration.

 \end{minipage}

 \end{table*}


%
%

\begin{table*}
\begin{minipage}{140mm}
\caption{Journal of observations. Run means the period in which observations were taken (as explained in the text). Filters and exp. time columns are in the same order, that for filters is: U, B, R, I,  H$\alpha$ and its cont., H$\beta$ and its cont. (for NGC6574, NGC6951 and NGC7714 no H$\alpha$ cont. image was available). For NGC5194 H$\beta$ images were taken but were not used afterwars due to questionable data values. These data have been removed from the table. Seeing was estimated using stellar images in each frame. Worst common seeing is tabulated, it has been computed to subtract the images as explained in the text. NGC1068, NGC3310, NGC7177 and NGC7469 were observed in a different run and are not included in this table  (\protect\citealt{dia:00}).}\label{journal}
 \begin{tabular}{@{}ccccccc}
  \bf Galaxy    & $\bf \alpha$     &$\bf \delta$      &\bf run           & \bf Filters                                                      & \bf exp. time (s)                                                   & \bf seeing      \\ 
 \bf                 &                      &                    &\bf                   &                                                                     &                                                                               & \bf FWHM ($^{\prime\prime}$)\\ 

 \hline
NGC 2782        & 09 14 05     & 40 06 49          & 5                    & U, B, R, I,                                                     &   1250, 800, 300, 600                                           &2.1                  \\
                        &                     &                        &                       &  H$\alpha$ (cont), H$\beta$ (cont)             &  2000 (2000),1800 (1800)                                      &                     \\
NGC 2903        & 09 32 10      & 21 30 02         & 3                    & U, B, R, I                                                      & 3600, 800, 300, 600                                             &  2.5               \\
                        &                     &                        &                       &H$\alpha$ (cont), H$\beta$ (cont)               & 1300 (2000), 3600 (1500)                                     &                         \\
NGC 3949        & 11 53 41      & 47 51 32         & 5                    & U, B, R, I                                                       &  2000, 100, 800, 500                                            & 3.6                  \\
                        &                     &                        &                       &H$\alpha$ (cont), H$\beta$ (cont)                & 2000 (2000), 2000 (2000)                                     &                         \\
NGC 3982        & 11 56 28      & 55 07 36         & 4                    & U, B, R, I                                                      & 1000, 800, 500, 700                                              & 2.1                   \\
                        &                     &                        &                       &H$\alpha$ (cont), H$\beta$ (cont)                & 1200 (1200), 1000 (1000)                                     &                            \\
NGC 4314        & 12 22 32      & 29 53 43         & 3                    & U, B, R, I                                                      &  1400, 1200, 1200, 450                                         & 2.0                \\
                        &                     &                        &                       &H$\alpha$ (cont), H$\beta$ (cont)                & 1500 (1500), 1200 (1200)                                     &                            \\
NGC 4321        & 12 22 55      & 15 49 20         & 4                    & U, B, R, I                                                       & 560, 80, 70, 60                                                     & 3.6                  \\
                        &                     &                        &                       &H$\alpha$ (cont), H$\beta$ (cont)                & 900 (900), 900 (900)                                             &                            \\
NGC 5033        & 13 13 28      & 36 35 38         & 4                    & U, B, R, I                                                       &  1000, 200, 300, 200                                             & 2.3                  \\
                        &                     &                        &                       &H$\alpha$ (cont), H$\beta$ (cont)                & 800 (900), 800 (900)                                             &                            \\
NGC 5194        & 13 29 52      & 47 11 54         & 4                    & U, B, R, I                                                       & 1000, 200, 300, 200                                              & 2.3                   \\
                        &                     &                        &                       &H$\alpha$ (cont),
                         & 800 (900)                                            &                            \\
NGC 5248        & 13 37 32      &08 53 01          & 4                    & U, B, R, I                                                        & 1230, 1800, 250, 250                                           & 2.3                     \\
                        &                     &                        &                       &H$\alpha$ (cont), H$\beta$ (cont)                & 2000 (1200), 1200 (1200)                                      &                            \\
NGC 5929/30   & 15 26 06      & 41 40 14         & 4                    & U, B, R, I                                                       & 2000, 700, 210, 120                                              & 2.5                    \\
                        &                     &                        &                       &H$\alpha$ (cont), H$\beta$ (cont)                & 1200 (2000), 1200 (1200)                                      &                            \\
NGC 5953/54   & 15 34 32     & 15 11 42         & 4                     & U, B, R, I                                                      & 1200, 1015, 250, 70                                               &  2.0                         \\
                        &                     &                        &                       &H$\alpha$ (cont),  H$\beta$ (cont)               & 1000 (1200), 1200 (1200)                                      &                            \\
NGC 6574        & 18 11 51      & 14 58 50         & 4                    & U, B, R, I                                                       & 1000, 400, 200, 50                                                &  1.8                         \\
                        &                     &                        &                       &H$\alpha$, H$\beta$ (cont)                           & 1000, 1000 (1000)                                                 &                            \\
NGC 6951        & 20 37 15       & 66 06 20        &4                     & U, B, R, I                                                       & 1500 (500), 300 (100)                                            &  1.9                         \\
                        &                     &                        &                       &H$\alpha$, H$\beta$ (cont)                           & 800 (800), 900 (1200)                                                 &                            \\
NGC 7714       & 23 36 14       & 02 09 19         &4                     & U, B, R, I                                                       &  900, 300, 200, 200                                              &           2.3 \\
                        &                     &                        &                       &H$\alpha$, H$\beta$ (cont)                           & 800 (900), 1200 (1200)                                                 &                            \\
 
\hline
\hline
 \end{tabular}
\end{minipage}

 \end{table*}


\section{The sample}

In order to achieve our scientific goals, we have studied a diverse population of galaxies with reported circumnuclear rings of star forming regions in the bibliography. Catalogues (Atlas of peculiar galaxies, \citealt{arp:66}) and different sources on circumnucear activity of galaxies were used to select the initial set of galaxies. See \cite{col:97} and \cite{gon:97} as an example. Among this larger sample of galaxies known to harbour circumnuclear structures, we chose those that best meet the criteria:
\protect
\begin{itemize}
  \item They represent as many nuclear types as possible.
  \item They represent as many Hubble types as possible.
  \item We have chosen interacting and non-interacting galaxies.
  \item They are ``local galaxies". We chose nearby galaxies to maximize the physical linear resolution.
  \item They could be observed close to the zenith in the appropriate epoch.
  \item High surface brightness and active nuclear star formation, were also taken into account.
  \item Face on galaxies were prioritized. A first visual selection of possible targets on published images of ring galaxies was made and ``no easily identified face on galaxies" were discarded. No attempt was made to correct linear quantities for inclination, since not for all galaxies the values of inclination were available from the literature. The data of major (D) and minor (d) diameters were obtained from NED for each galaxy. The quotient d/D  has been computed for each object as a simple estimation of inclination. It varies from 0.9, nearly face on objects, to 0.4 in NGC3949 and NGC5930, which are the most inclined.
 \end{itemize}

	The final sample consists of 20 mostly early-type spiral galaxies (four of them included also in \citealt{dia:00}) with distances between 8.4 and 65 Mpc (for Ho = 75  Km s$^{-1} Mpc^{-1}$); it includes barred (70\%) and non-barred (30\%) galaxies and present different nuclear activity levels: HII galaxies: 30\%; LINER: 15\%; Seyfert 2: 40\%; Seyfert 1: 10\%; no activity reported: 5\%. 
	Regarding the environment, 60\% of the objects are considered interacting galaxies. A detailed description of the sample can be seen in Table \ref{sample}.

Details of the observed galaxies are given in the Appendix.

\section{Observations and data reduction}
The data were acquired during five observing runs. For the first two runs (from 1988 to 1990), we used a blue sensitive GEC CCD at the f/15 Cassegrain focus of the 1.0-m. Jacobus Kaptein Telescope (JKT) of the Isaac Newton Group at the Observatorio del Roque de los Muchachos, La Palma, Spain. The CCD had 578x385 pixels 22 $\mu$m wide. The scale obtained with this instrumental configuration is 0.3 arcsec pixel $^{-1}$ and the CCD field of view is 2.89 by 1.92 arcmin$^{2}$.
Observations and reduction procedures for these runs are described in detail in \cite{dia:00}.

	The last three observing runs were carried on from 1999 to 2000 at the Centro Astron\'{o}mico Hispano Alem\'{a}n de Calar Alto, Almer\'{i}a, Spain. We used a chip Tektronics TK 1024 at the 1.52 m Spanish Telescope. This CCD had 1024 x1024 pixels with a geometrical scale of 0.4 arcsec pixel $^{-1}$ and a field of view of 6.9 by 6.9 arcmin$^{2}$.

	A journal of observations can be found in Table \ref{journal}, except for NGC1068, NGC3351, NGC7177 and NGC7469, reported already in \citealt{dia:00}. All observations were made under photometric conditions, and the seeing was estimated using stars present in each frame. At the beginning and end of each night, and between different galaxies (taking into account air masses) photometric (\citealt{lan:83}) and spectrophotometric (\citealt{mas:88}; \citealt{sto:77}) standard stars were observed to perform flux calibrations.\footnote[1]{Standard stars observed: HD205556, HD192281,HD217086, 95 52, 98 978, 94 308, 92 336, hilt600, feige25, feige15, hz15, feige34, feige66, G191B2B, hz44, pg17, hd192, wolf1346,100304, 52529, 104598, 1061024, 94308, 97503, 97507, g191, 1121242,113442, 209796, 188934, 14774}
		We also took dome and sky flat-field images as well as zero exposure time frames to set the bias level.\\ 
	The images were taken in the following order for each source: $H\alpha$, continuum $H\alpha$, $H\beta$, continuum $H\beta$, R, I, B and U. If possible two images were taken of each galaxy. As the runs covered several nights, the frames were checked the following day, and if any of them was faulty, it was rejected and the galaxy was observed again.

	The data were processed and analyzed using IRAF and MIDAS \footnote[2]{IRAF: the image reduction and analysis facility is distributed by the National Optical Astronomy Observatories which is operated by the Association of Universities for Research in Astronomy, (AURA) under cooperative agreement with the National Science Foundation. MIDAS is run by the European Southern Observatory (ESO).} routines including the usual procedures of removal of cosmic rays, bias subtraction, division by a normalized flat field and flux calibration.
Sky background correction was performed by averaging the mean count values in several boxes in the outer part of each frame. Cosmic rays were removed and the Point Spread Function (PSF) was estimated using stellar images in each frame.

	Images in the broad U, B, R, I, narrow H$\alpha$ and H$\beta$, and continuum filters (redshifted if necessary) were taken for each galaxy as described in the journal of observations (Table \ref{journal}). The narrow and broad filter characteristics are given in Table \ref{narrow-f}. Note that the H$\alpha$ filter includes the [NII] lines at $\lambda\lambda$ 6548, 6584 \AA, although their contribution to the filter is at about half transmission. The typical value of the [NII] $\lambda$ 6548 \AA, line intensity in HII regions is at most 1/2 of H$\alpha$. (\citealt{den:02}; \citealt{dia:07}; \citealt{dia:00})\\
	
	Taking into account that in very crowded regions with no constant background, measuring fluxes from diffuse object is quite difficult and to a certain degree some arbitrary decisions are made, the following criteria have been chosen to determine the HII region sizes and fluxes:
	In isolated regions, we have computed the  H$\alpha$ flux inside boxes of different sizes. Radius stands
for the equivalent radius of the circular region having the same area as the square one.
$r = \sqrt{Area/\pi}$. 

In the plot of  flux {\it vs}
radius \citep[see fig 2,][]{dia:00}, an asymptotic behaviour can be seen. For most observed 
regions the asymptotic radius is reached when the flux falls down to 10 \% 
of the flux at the central pixel. We have taken this radius as the radius of 
the region, and the flux inside it as the region flux.

When the morphology of the region is distorted \citep[see fig 2,][]{dia:00}, we have added all the flux inside a circular aperture up to the isophote corresponding to 10\% of the central value, even though the minimum could be lower.
When two regions are very close to each other we have computed their flux together if the minimum signal along the line that
joins the two central pixels is higher than 50\% of the smaller 
maximun. In all other cases, we have integrated both regions separately.
	
	Both atmospheric and galactic extinction corrections were applied. The first one using the extinction coefficients provided by La Palma Observatory.  For galactic extinction we used the values of E(B-V) taken from \cite{bus:84}; \cite{sch:98} and \cite{sea:79}.

H$\alpha$ and continuum images were aligned according to the offsets derived from the centroids of a bi-dimensional Gaussian fit to the field star images. Afterwards H$\alpha$ frames were continuum subtracted in order to obtain net H$\alpha$ line images.
If seeing values were significantly different between the H$\alpha$ and continuum frames, the best one was degraded to fit the worst image. If no continuum frame was available, an artificial one was constructed from broad-band images as explained in \cite{ter:91}. Discussion on the convenience and accuracy of this method can be found in \cite{dia:00}.
Fig. \ref{contours} shows the H$\alpha$+[NII] countinuum subtracted line contour maps for all observed galaxies. Individual HII regions are marked and labelled. In the images, north is to the top and east is to the left.

In order to measure the broad band fluxes, all the frames 
were carefully aligned as explained above, and the corresponding fluxes, for all filters, were 
measured inside the apertures defined on the H$\alpha$ line frame.  	
Subsequently, the broad band magnitudes were computed and these magnitudes 
were subtracted to get the colours.

\section{Analysis Methods}
\subsection{Limiting Radius}
In all studied galaxies the CNSFRs are arranged in a ring or pseudo-ring pattern (see Fig. \ref{contours}). The first question we had to answer at the very beginning of this work was: what does it mean circumnuclear? and the second: how close must a ring be to the galactic nucleus in order to be considered circumnuclear? 
It is easy to find in the literature the terms: outer, inner and nuclear applied to galactic rings in barred and non-barred galaxies; however, a common criterion to distinguish among these three types of rings, when just an only ring is observed in a single galaxy, has not been reached yet. 

When the three types of rings are observed in a given galaxy, the answer is easy: the innermost one is considered as circumnuclear. Something similar happens when a bar is present, because the bar usually fills in the inner ring, being responsible for the gas infall. But except in these two cases, rings are not generally classified in an unambiguous way.

\begin{figure}

\includegraphics[scale=0.37]{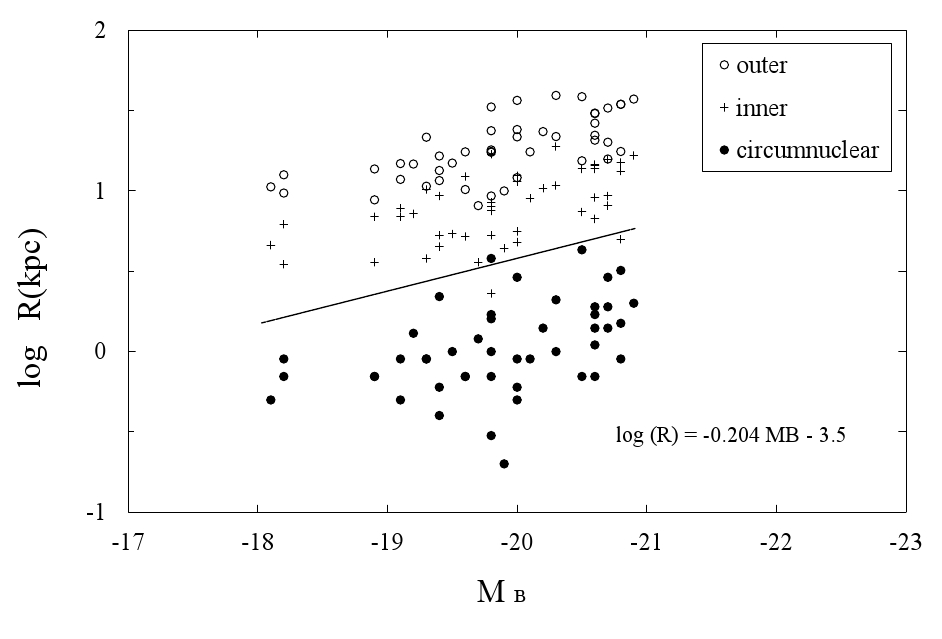}

\caption{The logarithm of the radii of different types of rings vs the absolute integrated magnitude of the host galaxy for the sample of  Buta \& Crocker (1993). For higher galaxy luminosities (lower integrated magnitudes), ring radii are also larger. This happens for nuclear, inner and outer rings. Nuclear rings seem to segregate occupying a region below the so called ``limiting radius''.} 
\label{limiting-radius}
\end{figure}

	Trying to solve this problem, we have developed the following procedure: from the sample of galaxies of \cite{but:93}, we have chosen those objects with three rings detected. No attempt has been made to correct for the inclination of the galaxy. We have made a logarithmic plot of the ring major axis in kpc vs the host galaxy absolute magnitude, M$_{B}$,  for each individual nuclear, inner and outer ring (see Fig. \ref{limiting-radius}).
Two facts can be observed: first, the larger the galaxy luminosity, the larger the radii of the rings. This happens for nuclear, inner and outer rings and, while a certain degree of overlapping exists between outer and inner rings, nuclear rings seem to behave in a slightly different way and a line can be visually drawn that separates the nuclear rings from the other two types. Points below that line, that we have called ``limiting radius line'' can be genuinely considered nuclear rings. 
This ``limiting radius'' for a given galaxy can be calculated as a function of the galaxy absolute magnitude as: :\\

log R(kpc) = -0.204 M$_{B}$ - 3.5\\ 

	Similar relationships can be found in the literature \citep[e.g.][]{kor:79} suggesting a dynamical origin for the rings. 
	Given the magnitude of the galaxy, the limiting radius can be simply computed, so HII regions beyond this radius will not be considered as circumnuclear. The limiting radius for each galaxy together with its integrated absolute Magnitude (B band) and some notes about ring morphology are given in Table \ref{lr}. Assuming that the ring lies in the galactic plane, the radius of the ring has been defined as the mean distance of each CNSFR to the galactic nucleus. The position of each CNSFR and the galactic nucleus corresponds to that of the pixel or pixels that show the maximum brightness in the  H$\alpha$ continuum corrected frames.

%
%

\subsection{HII region sizes and photometry}

From the H$\alpha$ continuum subtracted frames, the star forming regions were identified and their line fluxes measured.
Three main difficulties had to be confronted throughout this process:\\
1.- Measuring fluxes from diffuse objects in a non-constant background is not easy, even with most modern software packages. In this kind of galactic environments the assignment of a radius to each region is really difficult and often relies on the experience of the astronomer.\\
2.- In very crowded regions, deciding the limits of two adjacent regions is almost arbitrary.\\
Theoretically, a radiation bounded HII region can be characterized by its Str\"{o}mgrem radius but, in practice, it shows an inner bright core surrounded by a more extended and dimmer region. The frontier between both regions can only be well distinguished in isolated and non distorted regions. Taken the former into account we have computed H$\alpha$ fluxes for a total number of 336 HII regions following the procedure described below.\\
For isolated regions we have computed the H$\alpha$ flux inside circular apertures of different radii \footnote[2] {Due to software constraints no circular regions could be integrated interactively, so we used square boxes instead. The term radius then stands for the radius of the circular region having the same area as the square one.}(See fig. \ref{limiting-radius} of \cite{dia:00} for examples and  a more extended discussion). For most of the observed regions an asymptotic behaviour is reached at approximately 10\% of the flux of the central pixel. We have taken this radius as the region-radius, and the region-flux is therefore, the integrated flux inside this radius. This procedure has also been used even if the morphology of the region seems distorted.
When two regions are so close that the formerly described 10 per cent minimum cannot be reached, the flux of both regions was considered together (and so the radius) if the minimum signal along the line that joins the central pixels of both regions is higher than half of the smaller maximum. In all other cases, the two  regions have been considered as different, and their radii have been computed from the center of each region (maximum pixel) to that minimum.\\
Once the size and H$\alpha$ flux of every region were measured, H$\alpha$ equivalent widths (computed by division of the H$\alpha$ line frame by the corresponding continuum image) were also obtained.
In order to measure the broad-band fluxes, all the frames were carefully aligned as explained above, and corresponding fluxes were measured inside the apertures defined on the H$\alpha$ line frame. Broad-band magnitudes were computed and these magnitudes were subtracted to obtain colours.

\subsection{Extinction}
	Having narrow H$\alpha$ and H$\beta$ images allowed us to also take into account internal absorption (Table  \ref{narrow-results}). This fact is of great importance given that internal extinction and subsequent reddening affects line and continuum emission. Its knowledge is a cornerstone to disentangle age and metallicity effects. 
A$_V$ computed values (Table \ref{narrow-results}) have been obtained using the extinction curve given by \cite{sea:79}, assuming theoretical values for the Balmer decrement in case B recombination (\citealt{bro:71}). When no reliable narrow continuum value was available, R or B filters were considered instead to get a net H$\alpha$ or H$\beta$ image. This method was tested with images obtained with the standard procedure and no systematic errors were detected. For some galaxies A$_V$ was taken from the literature since no H$\beta$ or H$\alpha$ observations were available. References for those cases are:
\begin{itemize}
\item[-] For NGC 1068, \'{A}lvarez-\'{A}lvarez et al (2001)
\item[-] For NGC 3310, \cite{pas:93}
\item[-] For NGC 5194, \cite{hil:97}
\item[-] For NGC 7177 and NGC 7469, \cite{ken:83}
\end{itemize}

	In some cases, and only for U-B colours, the A$_V$ value obtained using Seaton´s curve, did not match what is expected for young O-type stars. This effect is not due to the presence of [NII] lines in the H$\alpha$ narrow filter as checked in \cite{alv:01} using spectrophotometric data from NGC 1068. In these cases the extinction curve proposed by \cite{cal:94} has been used instead.

\subsection{Characterizations of uncertainties}
Three  main sources of error have been taken into account:\\
1.- Non uniform background.\\
2.- Narrow line contaminating broad-band filters.\\
3.- Radius estimation.\\

	To estimate the background contribution to the computed magnitudes, we have used the intensity radial profiles given in \cite{sch:00} for the galaxies  in common with our sample. The mean contribution is about 10\% in the I band, so a mean error of about 0.03 magnitudes in colours should be considered.\\
	Bluer filters are less affected by background contribution, so this value might be considered as an upper limit

%
%

\begin{figure}
\includegraphics[scale=0.34]{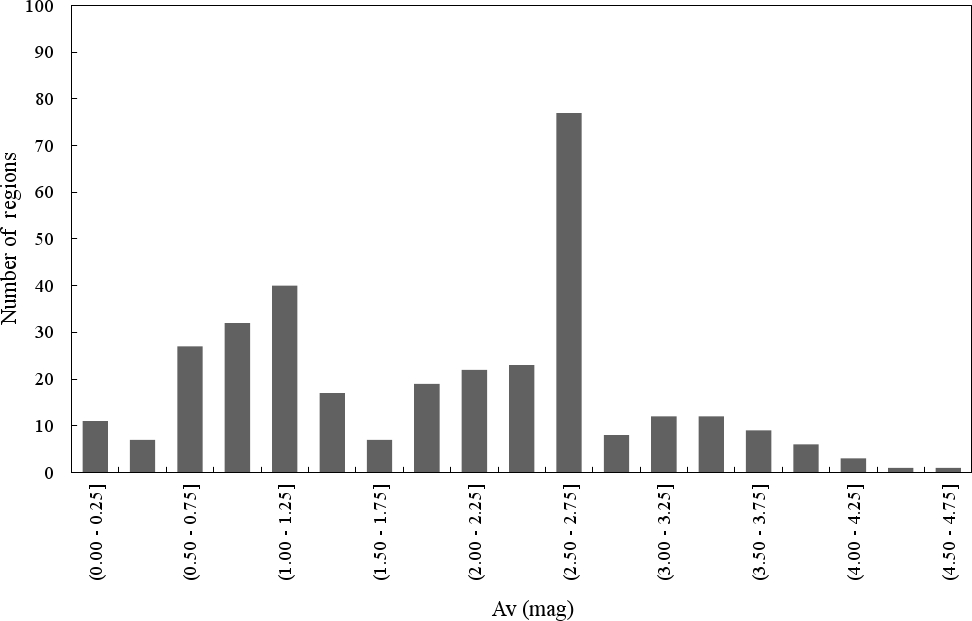} 
\caption{Distribution of the visual extinction in magnitudes for the observed regions in histogram form. The bin at the 2.5-3.75 magnitude range includes  62 regions in NGC~5194 (M51) for which no $H\beta$ line observations were secured. Data from far-UV, U, $H\alpha$, and R images by Hill et al. (1997) were used instead. Although in a sense this maximum may be considered an artifact, it has been included for comparative purposes.} 
\label{extinction}
\end{figure}

%
%

\begin{table*}
 \begin{minipage}{150mm}
 \caption{Filter characteristics}\label{narrow-f}
 \begin{tabular}{@{}ccc}
\bf NARROW FILTER  $\bf \lambda(\AA)$ & \bf z(km/s) & $\bf  \bigtriangleup \lambda(\AA)$ \\
          \hline
  6563 ($H\alpha$)              &0                   &53     \\

  6607                               &2000            &50       \\

  6700                               &6000            &50        \\

  4869 ($H\beta$)                &0                   &112       \\

  4695 (v Str.)                   &0                   &188        \\

 \hline
 \end{tabular}
 \begin{tabular}{@{}cccc}
\bf BROAD BAND FILTER                            &$ \bf \lambda(\AA)$ & $\bf \bigtriangleup \lambda(\AA)) $& \bf  max. transmission \\
          \hline
U                                       &3500                     &1000                                                  &60\%   \\

B                                       &4350                     &2000                                                   &65\% \\

R                                       &6750                     &2500                                                   &80\%    \\

I                                          &8000                   &2000                                                    &90\% \\

 \hline
 \end{tabular}
\end{minipage}

 \end{table*}


%
%

\begin{table*}
 \begin{minipage}{170mm}
 \caption{Ring Morphology and computed limiting radius on continuum subtracted H$\alpha$ images as described in the text.}\label{lr}
 \begin{tabular}{@{}ccccl}
  \bf Galaxy & $\bf M_{B}$ & \bf Limiting             & \bf Number      &\bf Ring Morphology  \\
                  &                   & \bf     Radius (Kpc) &  \bf of regions  &                         \\
 \hline
  NGC 1068 &-21.7     &  3.7    &        25    &    Closed ring. Highest emission following  [OIII].  \\
  NGC 2782 & -21.7        & 3.7   &        6     &  Closed partially resolved ring. Region 5, possibly a star? \\
  NGC 2903  & -20.0    &  2.2    &           21 &  Star forming activity is widespread throughout the circumnuclear region. \\
  NGC 3310  & -19.3    &   1.8   & 21           &    Closed ring. Nucleus off centre.\\
  NGC 3949  & -19.9    & 2.1     &   24         &    Pseudo-spiral structure inside patchy ring.  \\
  NGC 3982  &  -19.1   & 1.7     &     42       &    Well organized closed ring. \\
  NGC 4314  & -18.5    & 1.4     &     9       &     Closed compact ring. Pseudo-spiral structure inside the ring.\\
  NGC 4321  & -20.7    &  2.7    &    10        &     Compact ring. High extinction in NW regions. \\
  NGC 5033  &  -20.7   &   2.7   &     38       &     Broken ring to the W, due possibly to high extinction.\\
  NGC 5194  &  -21.0   &   3.0   &     62       &      Double nucleus or region very close to it. Inner star formation.    \\
  NGC 5248  &  -20.0   &  2.2    &     14       &      Widespread star formation structures in inner region\\
  NGC 5929  & -18.6    &  1.4    &          5  &        Patchy ring\\
  NGC 5930  &  -19.1   &   1.7   &        3    &        Patchy ring\\
  NGC 5953  & -18.8    &   1.5   &      5      &       Symmetric closed ring\\
  NGC 5954  &  -18.4   &   1.3   &     5       &        Irregular ring. Highly inclined galaxy\\
  NGC 6574  &  -19.7   &    2.0  &       11     &      Symmetric ring, broken to the NE.   \\
  NGC 6951  &  -20.3   &    2.4  &      6      &       Closed and compact ring. Inner star formation.\\
  NGC 7177  &  -18.9   &    1.6  &    11      &      Nuclear star formation with two big HII regions at the base of spiral arm.\\
  NGC 7469  &   -21.1  &     3.1 &      11      &     Patchy ring only partially resolved.  \\
  NGC 7714  &   -21.3  &     3.3 &         7   &      Patchy ring, open to the W. HII regions very close to the nucleus. \\

 \hline
 \end{tabular}
  \end{minipage}
 \end{table*}

	The most prominent lines present in the broad-band filters are:\\
U: [OII] 3727 \AA; [NeIII] 3868 \AA.\\
B: H$\beta$; [OIII] 4959, 5007 \AA; [NI] 5194 \AA.\\
R: [OI] 6300, 6363 \AA; [SII] 6717, 6731\AA; H$\alpha$+[NII] 6548,6584 \AA.\\
I: [SIII] 9069,9532 \AA\\
In order to check the effect that these lines could have on the observed colours and therefore on the  conclusions of this work, we have added narrow lines synthesized by \cite{sta:96} to Starburst99 continuum emission models for a single burst of star formation.
We have chosen physical parameters for the gas that match those most commonly observed in star forming regions located in the very inner parts of galaxies, namely: solar or oversolar abundances, Mup=100 M$_\odot$, M$_\star$=$10^{6}$ M$_\odot$, n$_{e}=10 cm^{-3}$.
H$\alpha$ and [OII] lines reach their maximum effect at very early ages of the burst (less than 3 Myr), while H$\beta$ and [OIII] do it later.  Around their maximum, between 15 and 20\% of the theoretical flux in the corresponding broad band has its origin in the narrow lines. These theoretical values hence represent an upper limit given that no additional, non-ionizing star population has been considered. When compared with observed values, the H$\beta$ flux amounts to about 6\% of the B filter at its maximum. 
Although the presence of narrow lines has no much influence on colours, the R and B filter fluxes have been corrected for H$\alpha$ and H$\beta$ emission.\\

	Despite all the above, the main source of uncertainty comes from the right determination of the radius of each HII region (see discussion in Section 4.2). Therefore, in order to estimate realistic errors, the fluxes inside the adopted radius plus and minus one pixel have been computed for each region in each frame. The absolute value of this error is bigger that that associated with narrow line contamination or non uniform background and therefore is the only error value quoted.It has been included in Table 5 as $\triangle$ F and in Table 6.

%
%

\begin{figure*}
\includegraphics[scale=0.5]{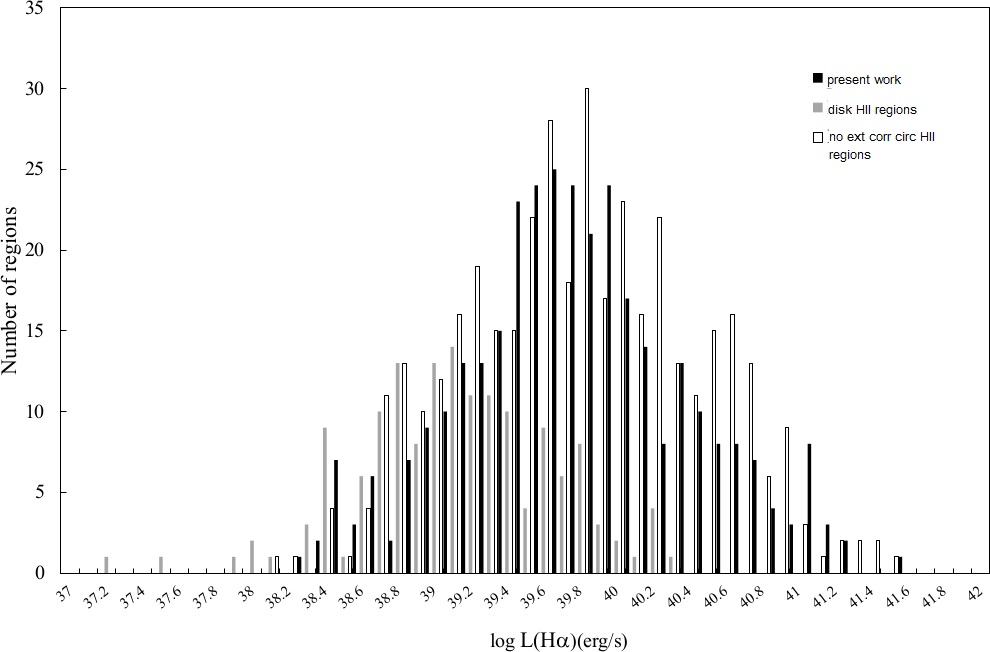} 

\caption{$H\alpha$ extiction corrected values histogram. Mayya \& Prabhu (1996) values for disk HII region and Mazzuca et al. (2008) (not corrected for internal extinction) values for cirmnuclear regions have been included for comparative purposes.} 
\label{Halpha-lum}
\end{figure*}


\section{Results and discussion}
The results of our measurements for each of the studied regions in a given galaxy can be found in Tables \ref{narrow-results} and \ref{broad-results}. Table \ref{narrow-results} lists in column 1 the identification number of the region; in column 2 the distance for the galaxy nucleus in arcsec; in column 3 the position from the galaxy nucleus in arcsec; in column 4 the logarithm of the H$\alpha$ luminosity in units of erg s$^{-1}$; in column 5 the quotient $\triangle$F over F; in column 6 the equivalent width of the H$\alpha$ line in \AA ; in column 7 the equivalent width of the H$\alpha$ line in \AA ; and finaly in column 8 the value of the visual extinction A$_V$ in magnitudes.  For those objects with no $H\alpha$ or $H\beta$ available, published photometric or spectrophotometric values have been used. See section 2 for references on individual objects.

Table \ref{broad-results} lists, for each individual galaxy, in column 1 the identification number of the region; in column 2 the radius of the region in pc (as defined in section 2.1); in column 3 the  distance from the galaxy nucleus in kpc; in column 4 the absolute magnitude in the R band; and in columns 5, 6 and 7 the U-B, B-I and R-I colours respectively. 

%
%

\begin{table*}
 \begin{minipage}{130mm}
 \caption{Narrow band results. No correction has been made for the inclination of the galaxy as explained in the text. Average seeing values were 2 (arcsec FWHM). $\frac{\triangle Flux}{Flux}$ is a measure of error in the flux calculation as explained in the text.  Only NGC~2903 is shown as an example. Both tables 5 and 6 are available in electronic format (malvar4\makeatletter@roble.pntic.mec.es). }
\label{narrow-results}
 \begin{tabular}{@{}cccccccc}
  \bf Region    & \bf Radius of        &\bf Position     from               &\bf Log L(H$\alpha$)     &\bf $\displaystyle\frac{\triangle Flux}{Flux}$         & \bf EW H$\alpha$  (\AA)  & \bf EW H$\beta$(\AA)  &\bf Av(mag) \\
   \bf number   &\bf region                & \bf   nucleus              & \bf erg s$^{-1}$                    &            &           & \\
   \bf   &\bf           (arcsec)       & \bf   (arcsecxarcsec)              & \bf                     &            &           & \\
\hline

\hline
&&&NGC 2903&	&&\\						
\hline						
1&4.5&(0.1, 0.1)&41.0&0.08&61.8&27.6&2.83\\
2&2.5&(0.1, -6.4)&40.1&0.23&54.4&20.6&1.70\\
3&1.2&(1.9, -12.2)&38.5&0.56&49.8&16.2&0.76\\
4&8.6&(18.9, -26.7)&39.8&0.02&26.0&26.2&0.76\\
5&1.6&(31.9, -28.6)&39.2&0.37&90.9&10.6&2.73\\
6&3.7&(34.7, -32.0)&40.1&0.11&84.0&16.2&3.23\\
7&1.2&(-8.5, -9.8)&38.4&0.59&61.9&25.1&0.60\\
8&4.1&(-19.1, -5.1)&39.8&0.13&48.7&12.7&2.47\\
9&4.9&(-22.7, 7.6)&39.5&0.09&40.2&18.1&1.16\\
10&3.3&(-14.2, 6.8)&38.5&0.12&17.6&14.7&0.80\\
11&1.6&(-31.1, 3.0)&38.9&0.36&68.0&9.3&2.18\\
12&3.3&(-27.8, 13.4)&39.7&0.19&65.7&13.4&2.57\\
13&3.3&(-32.0, 17.4)&39.5&0.19&66.7&13.9&2.08\\
14&5.7&(-35.0, 24.6)&40.1&0.07&59.5&18.1&2.40\\
15&2.1&(-36.0, 32.0)&39.8&0.28&117.0&9.9&3.83\\
16&3.7&(-33.4, 36.0)&40.0&0.12&76.3&14.7&2.87\\
17&7.8&(-35.1, 36.1)&39.9&0.06&38.5&20.2&1.56\\
18&2.5&(-12.9, 19.8)&38.5&0.22&22.9&13.6&0.87\\
19&3.7&(29.1, 23.0)&40.1&0.08&95.3&14.0&3.47\\
20&1.2&(23.5, 23.7)&38.7&0.59&83.5&7.3&2.59\\
21&2.9&(26.8, 37.6)&39.1&0.12&53.7&11.6&2.02\\
\hline
 \end{tabular}               
 \end{minipage}
 \end{table*}

%
%

\begin{table*}
 \begin{minipage}{140mm}
 \caption{Broad band results. Linear radius of each region is affected by a mean error due to: seeing conditions, size estimation, etc. of about 20\% on average. Error in $M_{R}$ and colors have been computed using radius of +1 and -1 pixels as explained in the text.}
 \label{broad-results}
 \begin{tabular}{@{}ccccccccccc}
  \bf Region    & \bf Radius of the        &\bf Distance from               &\bf M$_R$            &\bf  $\pm $ &\bf  (U-B)  &\bf  $\pm$ & \bf (B-R)  &\bf  $\pm$  &\bf  (R-I)  &\bf  $\pm $\\
   \bf number   &\bf  region (pc)            & \bf   nucleus (Kpc)           & \bf                  &            &                &           &                 &            &                 & \\
\hline
&&&NGC 2903&&&&&&&										\\
\hline									
1  &  183$\pm$37  &  0.007  &  -18.2  &  0.2  &  -0.88  &  0.01  &  -0.20  &  0.00  &  0.46  &  0.00\\
2  &  100$\pm$20  &  0.262  &  -16.4  &  0.3  &  -0.54  &  0.01  &  0.17  &  0.07  &  0.68  &  0.03\\
3  &  50$\pm$10 &  0.501  &  -13.7  &  0.7  &  -0.16  &  0.01  &  0.80  &  0.05  &  0.79  &  0.01\\
4  &  350$\pm$70  &  1.330  &  -16.6  &  0.1  &  0.05  &  0.00  &  1.10  &  0.01  &  0.99  &  0.00\\
5  &  66$\pm$13  &  1.742  &  -13.7  &  0.5  &  -0.85  &  0.03  &  -0.07  &  0.04  &  0.69  &  0.03\\
6  &  150$\pm$30  &  1.920  &  -15.5  &  0.2  &  -0.94  &  0.03  &  -0.25  &  0.01  &  0.57  &  0.01\\
7  &  50$\pm$10 &  0.527  &  -12.9  &  0.8  &  -0.33  &  0.00  &  0.80  &  0.04  &  0.88  &  0.02\\
8  &  166$\pm$33 &  0.806  &  -15.6  &  0.2  &  -0.80  &  0.25  &  0.08  &  0.29  &  0.74  &  0.24\\
9  &  200$\pm$40  &  0.974  &  -15.4  &  0.2  &  -0.49  &  0.04  &  0.59  &  0.03  &  0.88  &  0.02\\
10  &  133$\pm$27  &  0.640  &  -14.7  &  0.2  &  -0.07  &  0.01  &  1.09  &  0.01  &  1.02  &  0.00\\
11  &  66$\pm$13  &  1.272  &  -13.3  &  0.5  &  -0.80  &  0.07  &  0.18  &  0.03  &  0.72  &  0.01\\
12  &  133$\pm$27  &  1.254  &  -15.0  &  0.3  &  -0.78  &  0.01  &  0.04  &  0.00  &  0.71  &  0.01\\
13  &  133$\pm$27  &  1.481  &  -14.9  &  0.3  &  -0.74  &  0.02  &  0.11  &  0.00  &  0.72  &  0.01\\
14  &  233$\pm$47 &  1.738  &  -16.0  &  0.1  &  -0.79  &  0.02  &  0.05  &  0.01  &  0.69  &  0.00\\
15  &  83$\pm$17  &  1.958  &  -14.5  &  0.4  &  -1.15  &  0.00  &  -0.55  &  0.01  &  0.52  &  0.03\\
16  &  150$\pm$30  &  1.998  &  -15.6  &  0.2  &  -0.86  &  0.03  &  -0.17  &  0.01  &  0.58  &  0.02\\
17  &  316$\pm$63  &  2.047  &  -16.3  &  0.1  &  -0.43  &  0.01  &  0.51  &  0.01  &  0.86  &  0.00\\
18  &  100$\pm$20  &  0.959  &  -14.3  &  0.3  &  -0.23  &  0.00  &  0.79  &  0.01  &  0.92  &  0.01\\
19  &  150$\pm$30  &  1.510  &  -15.5  &  0.2  &  -1.02  &  0.04  &  -0.41  &  0.01  &  0.53  &  0.01\\
20  &  50$\pm$10  &  1.357  &  -12.6  &  0.8  &  -0.74  &  0.06  &  0.12  &  0.07  &  0.75  &  0.02\\
21  &  116$\pm$33  &  1.880  &  -14.0  &  0.3  &  -0.56  &  0.04  &  0.29  &  0.01  &  1.00  &  0.01\\
\hline	

\hline
\hline

 \end{tabular}
\end{minipage}
 \end{table*}

Here we present the results for NGC 2903; for the rest of the sample, the tables and images are available, upon request at (malvar1@ing.uc3m.es)

\subsection{Extinction and reddening}

Extinction depends on distance to the source, wavelength, dust spatial distribution or even on time (\citealt{var:13}). A ``classic'' HII region (photoionized plasma, optically thin case B emission, electron temperature of 10000 K and electron density $\sim 10^{2} cm^{-3}$) has a theoretical value of $\frac{H\alpha}{H\beta}$ of 2.86 (\citealt{ost:89}; \citealt{bro:71}), therefore an observed $\frac{H\alpha}{H\beta}$ ratio significantly higher than 2.86 is evidence for reddening by dust. Some authors have reported that extinction of HII regions, derived from the comparison of Balmer lines together with radio continuum in the Magellanic Clouds and other external galaxies, is slightly higher than that derived from Balmer lines alone, so in a conservative sense, our values should be taken as lower limits (\citealt{mel:79}; \citealt{isr:80}).

Fig \ref{extinction} shows the distribution of the visual extinction in magnitudes for the observed regions in histogram form. The bin at the 2.5-3.75 magnitude range includes  62 regions in NGC~5194 (M51) for which no $H\beta$ line observations were secured. Data from far-UV, U, $H\alpha$, and R images by Hill et al. (1997) were used instead. Although in a sense this maximum may be considered an artifact, it has been included for comparative purposes.

We have found an A$_V$ average value of 1.85 (see Fig \ref{extinction}) with values ranging from 0 to 4.75 magnitudes for the most obscured regions. We do not find any correlation with galactocentric distance or with region size. The most luminous, but not the biggest, HII regions seem to show higher A$_V$ values. This effect,  bright compact sources showing high A$_V$ values, was also found by \cite{cap:96}. We also find a loose correlation between A$_V$ and colours, with redder regions showing higher extinction values. To find out if this effect is real, or if it is due to age or metallicity, a deeper study focusing only on extinction and reddening is needed. 

	Even though not many A$_V$ references on  individual external HII regions can be found in the literature, specially using the $\frac{H\alpha}{H\beta}$ ratio (Balmer fluxes are not always measured on the same source, or from exactly the same region, or they are not integrated over equivalent base-levels), our values are in good agreement with other published ones (\citealt{hob:97}; \citealt{cap:96}; \citealt{may:96}), for similar regions.

\subsection{Photometry results}

Fig \ref{Halpha-lum} shows a histogram of the extinction corrected H$\alpha$ luminosity of the studied regions.
H$\alpha$ luminosity values vary from 1.3 $10^{38}$ to 4 $10^{41 } erg s^{-1}$. Region 1 in NGC~5954 and  region 17 in NGC~3949 lay at the extremes of the luminosity range; 291 of the 336 regions observed show H$\alpha$ luminosities higher than log L(H$\alpha$) $>$ 39, hence they can be classified as supergiant HII regions (\citealt{knn:89}).
No clear correlation seems to exist between H$\alpha$ luminosity of the regions and interaction signatures of the host galaxy, nevertheless the most luminous HII regions are found in non interacting galaxies, with nuclear activity (Sy2) and Sb-Sbc morphologies. However, only two galaxies in our sample are classified as Sy2 and just one is non-active. These small numbers must be taken into account when reaching conclusions.

\begin{figure}
\includegraphics[scale=0.35]{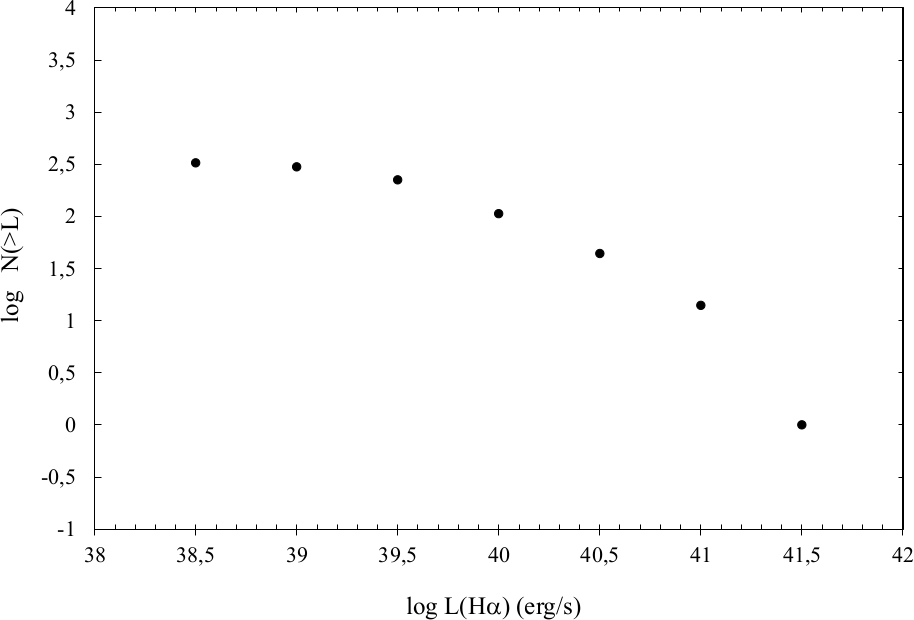} 
\caption{$H\alpha$ luminosity function. Note the break in the slope al high values of $H\alpha$ luminosity.} 
\label{lum-func}
\end{figure}

	An H$\alpha$ luminosity function such as:$\frac{dN(L)} {dL} \propto L^{-\alpha}$, has been considered, and the cumulative H$\alpha$ luminosity function
{$logN(> L) \propto (1+\alpha) log(L)$} has been computed for each galaxy, finding values for the slope that range from 1.22 to 2.88 in good agreement with those found by \cite{knn:89} and \cite{roz:96}.
This parameter seems to be related with the nature of the HII regions, and some authors (\citealt{roz:96}; \citealt{kna:98}; \citealt{bec:00}; \citealt{bec:02}  and \citealt{bec:96}) find a break between radiation and matter bounded regions at log (L(H$\alpha$)) of about 38.5. Given that we do not find this break, and being the luminosities of our studied regions all above this value (some small and low luminosity regions might have been missed due to overcrowding or resolution limits), we conclude that most of our CNSFR should be in the high luminosity matter bounded tail of the luminosity function. No other significant difference with normal disk HII regions has been found. Plotting the whole set of HII regions together (Fig \ref{lum-func}), we find that the luminosity function becomes  steeper at higher $H\alpha$ luminosities. Bins of 0.5 in log(L($H\alpha$)) have been considered given that this is the uncertainty of the log(L($H\alpha$))  computed for each region in a radius of plus and minus one pixel (See section 4.4).
The slope varies from 0.2 for less luminous regions, to 2.3 for the high luminosity tail. 
No correlation with nuclear activity or host galaxy interaction features has been found either.

\begin{figure*}
\includegraphics[scale=0.53]{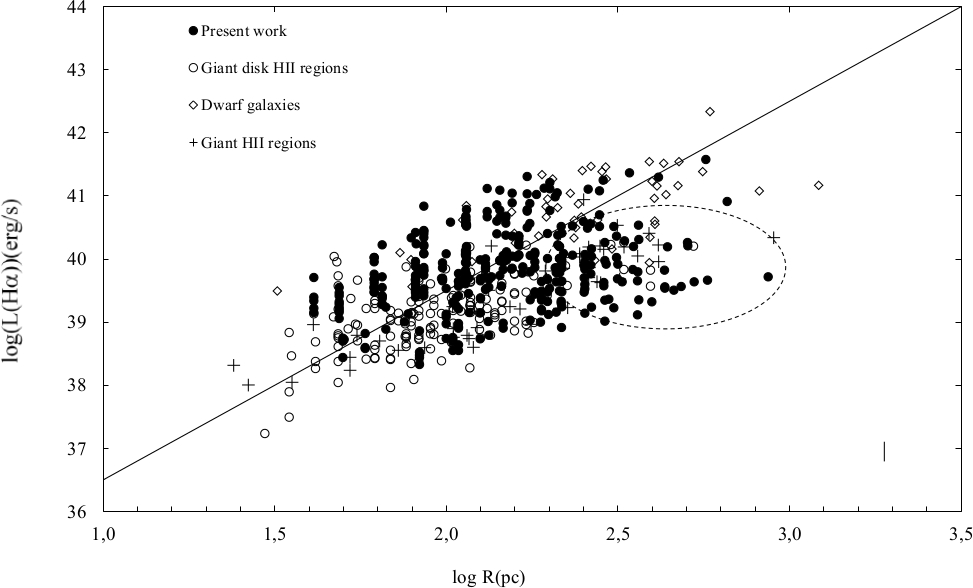} 
\caption{Present work regions compared with those observed by \protect \cite{may:94} for giant disk HII regions; \protect \cite{pop:00} for dwarf galaxies and \protect \cite{ars:88} for giant HII regions. The straight line represents slope 3 (radiation bounded case). Note the big regions inside the boundary oval dotted line which show luminosities smaller than those corresponding to radiation bounded regions of the same size. Maximum error bar is at the bottom right hand corner} 
\label{lum-rad}
\end{figure*}

Furthermore, radiation bounded regions follow a relationship between log[L(H$\alpha$] vs log(Radius) with a slope of about 3 (Fig. \ref{lum-rad}) (\citealt{ken:89}; \citealt{roz:96}). We have found for the individual galaxies of our sample an average value of 2.04 $\pm$ 0.6  with minimum and maximum values of 1.15 in NGC~5248 and about 3 in NGC~5954 (3.1), NGC~3949 (2.9) and NGC~5953 (3.2).
NGC~3949 also shows a slope break at log[L(H$\alpha$)] $\simeq$ 39.8. This may point to the HII regions in these latter three galaxies being radiation bounded. 

	Two main effects could be affecting the value of the slope making it appear lower; firstly, 
in overcrowded fields, the radius of a region can be underestimated with respect to isolated regions. This seems to happen more frequently for low luminosity regions. In this case the effect will be ``rising'' the line at small luminosities.  Secondly, bigger regions seem to be less luminous and more diffuse (see regions inside the dotted line in Fig. \ref{lum-rad}). These big, but otherwise low luminosity, regions may be matter bounded  (\citealt{roz:96}).

%
%

%
%

The equivalent width (EW) of the H$\alpha$ line is a good indicator for age and metallicity. The values that we have found range from 2.6 \AA\ in NGC~4314 (region 8) to over 900 \AA\ in NGC~3310 (region 21). About 25 \% of the regions studied show EWs smaller than 50 \AA; nearly 50 \% have values that range from 50 to 100 \AA~  and only 10 \% show values higher than 250 \AA.
This high EW tail is mainly conformed by HII regions located in galaxies with Sy2 or LINER nucleus. Values of EW higher that 450 \AA, that would imply ages younger than 6.5 Myr regardless of metallicity, are only found in interacting galaxies (Fig \ref{ewhist}).

\begin{figure}
\includegraphics[scale=0.35]{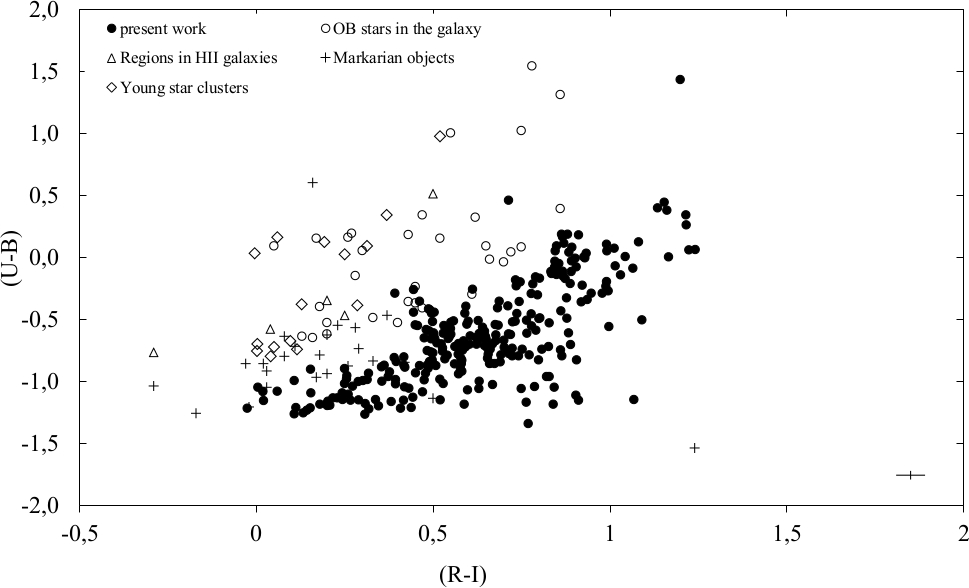} 
\caption{Colour-colour (U-B vs R-I) for circumnuclear (present work) HII regions, OB stars in the galaxy (\protect \citealt{lah:87}) regions in HII galaxies (\protect \citealt{igl:88}) young stars clusters (\protect \citealt{pat:01}) and Markarian objects (\protect \citealt{cai:01}). Average error bars are plotted at the bottom right hand corner.} 
\label{colour-colour}
\end{figure}

The continuum emission in the U and B filters comes mainly from young O and B stars ionizing the nebula, and to a lesser extent, from WR stars when present. \cite{men:00} found U-B mean colours of -0.76 for a sample of WR galaxies, and O, B stars show U-B colours ranging from -0.7 to 1.5 for solar metallicity (\citealt{lah:87}), and from -2 to 1 in under solar metallicity objects (\citealt{par:93}).
In our sample we find values of U-B that vary from -1.4 to 0.6 (Fig \ref{colour-colour}) which will imply burst ages ranging from 2 to 20 Myrs for over-solar metallicity models, following a bimodal distribution with two maximum values, -0.7 and -0.3 

\begin{figure}
\includegraphics[scale=0.35]{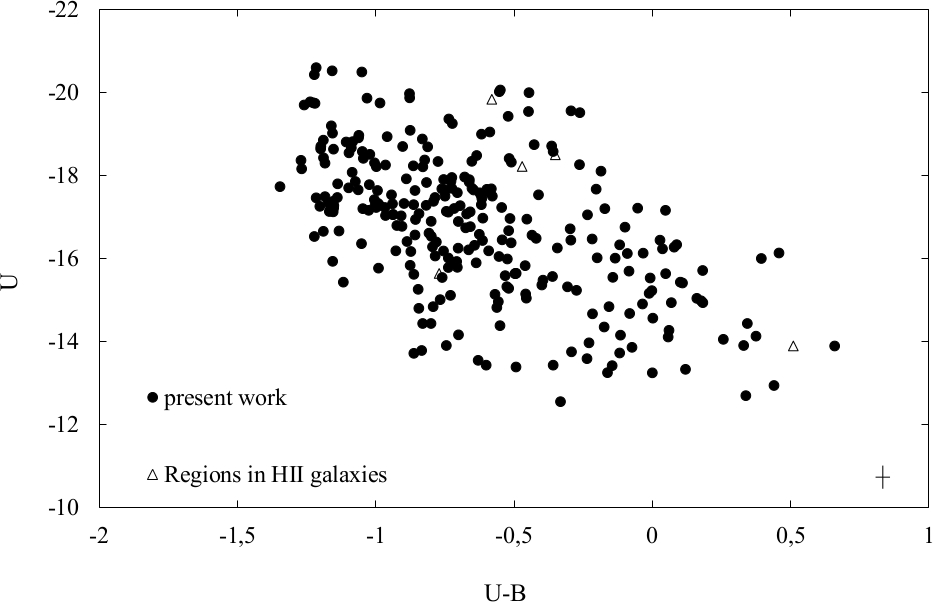} 
\caption{Colour-magnitude diagram in the blue for circumnuclear (present work) HII regions. \protect \cite{igl:88} data have been included for comparison purposes. Average error bars are plotted at the bottom right hand corner.} \label{u-u-b}
\end{figure}

 This second value mainly corresponds to HII regions located in non-interacting galaxies.
No clear trend is found in relation with morphological, nuclear type or barred structures. The colour-magnitude diagram [(U-B) vs U] (Fig \ref{u-u-b})
suggests a correlation in the sense that more luminous regions are also bluer, although some regions do not follow this trend.
Two effects seem to overlap: age -- older regions are less luminous and redder due to the death of massive OB stars -- and a mass effect, given that in some galaxies all their HII regions have constant U-B colour and a great dispersion in U values.

Continuum emission in the red has its origin mostly in stars in their red supergiant evolutionary phase which show an average R-I of 0.9. 
The colours that we observe vary between -0.02 and 1.22 for high metallicity objects, and from -0.12 to 0.16 for NGC 3310, the lowest metal content galaxy in our sample (Fig \ref{i-ri}), which match with bursts younger than 10 Myrs.
As in the U-B case, the distribution of R-I colour is bimodal with a first maximum at R-I $\sim$ 0.9, matching with non-interacting barred galaxies, and a second maximum at $\sim$ 0.6.
Higher dispersion values are found when compared with the U-B distribution, maybe due to a mixture of infrared sources.
Summing up, in the U vs U-B plot (Fig \ref{u-u-b}), HII regions show an homogeneous behaviour and values close to those of a young open cluster. This homogeneity does not exist in the I vs R-I color-magnitude diagram where metallicity and evolutionary effects can be seen. In both colour-magnitude plots, bluer regions seem to appear in galaxies with close companions and clearly seen interacting features and/or LINER or Seyfert 2 nucleus. By contrast, redder regions are more frequently found in isolated galaxies with HII or Seyfert 1 nucleus and clearly detected bars.
Differences among individual HII regions can be best detected in a colour-colour diagram. An overall correlation appears between both colours. Galaxies with  the bluest HII regions are: NGC 3982 (Sy2), NGC 5954 (Sy2), NGC 5953 (Sy2) and NGC 5194 (LINER). In the reddest part of the plot are located the following galaxies: NGC 4321 (HII), NGC 2782 (HII), NGC 5953 (HII) and NGC 5033 (Sy1).

%
%

%
%

%
%

\begin{figure}
\includegraphics[scale=0.35]{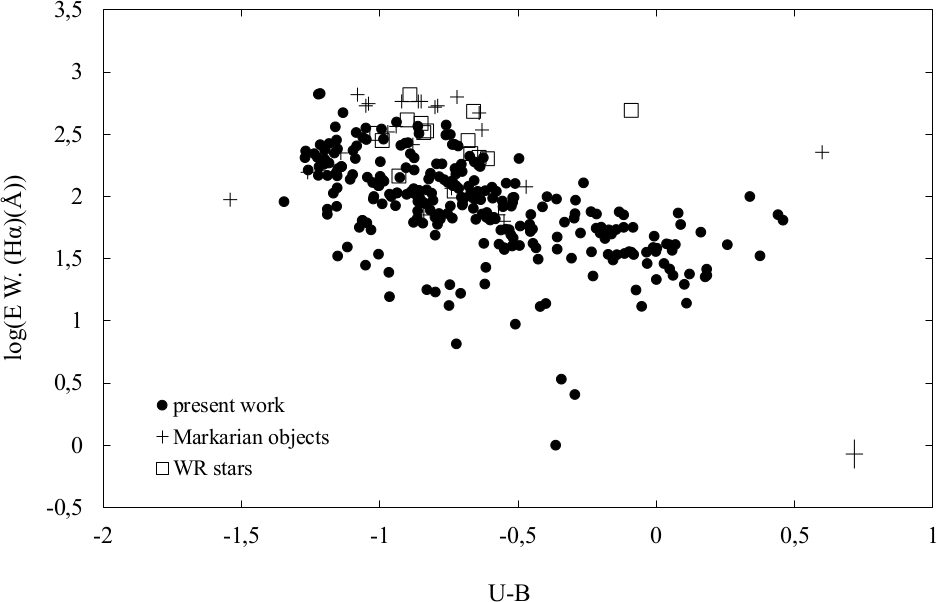} 
\caption{log EW($H\alpha$) vs U-B for circumnuclear  HII regions (present work),  WR stars (\protect \citealt{men:00}) and Markarian objects (\protect \citealt{cai:01}). Average error bars are plotted at the bottom right hand corner.} 
\label{EW-blue}
\end{figure}


%
%

\begin{figure}
\includegraphics[scale=0.35]{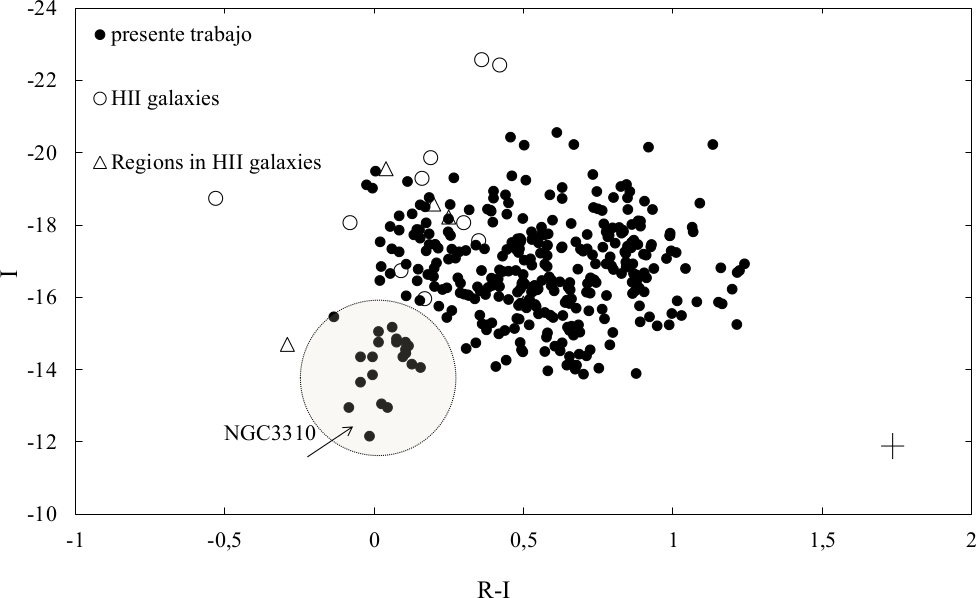} 
\caption{Colour-magnitude diagram in the red for circumnuclear HII regions  (present work), regions in HII galaxies (\protect \cite{igl:88}) and HII galaxies (\protect\citealt{ter:91}). Note the low metalicity effect in NGC3310, see \protect\cite{pas:93} where metallicity is obtained for some of the regions. Average error bars are plotted at the bottom right hand corner.} \label{i-ri}

\end{figure}

%
%

\begin{figure}
\includegraphics[scale=0.35]{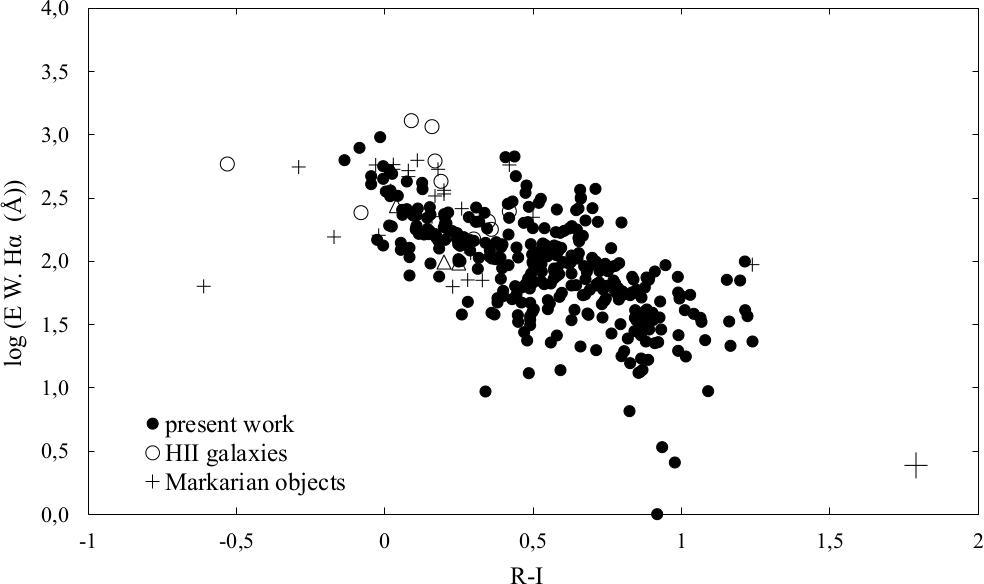} 
\caption{log EW($H\alpha$) vs R-I for circumnuclear (present work) HII regions, regions in HII galaxies (\protect \citealt{igl:88}), HII galaxies (\protect\citealt{ter:91}) and Markarian objects (\protect \citealt{cai:01}). Average error bars are plotted at the bottom right hand corner.} 
\label{EW-red}
\end{figure}

%
%

\begin{figure}
\includegraphics[scale=0.35]{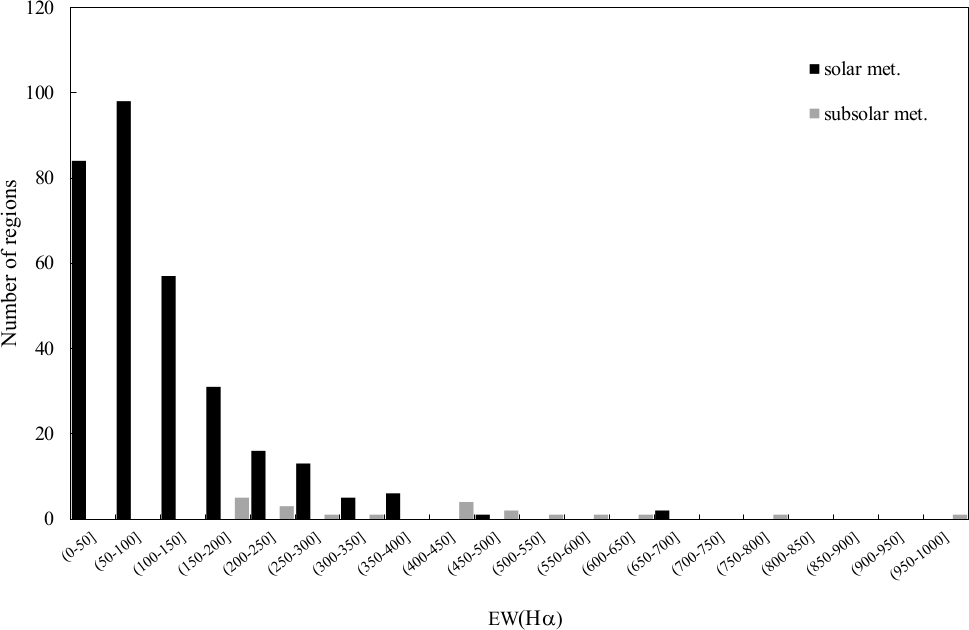} 
\caption{$H\alpha$ Equivalent Width (\AA) histogram for solar and subsolar (NGC 3310) luminosity rings. } 
\label{ewhist}
\end{figure}

Summing up the information provided by narrow and broad band filters together, some interesting facts can be pointed out: i) The bluest regions appear mainly in interacting galaxies with LINER or Sy2 type nucleus. These regions are also strong $H\alpha$ emitters and show the highest values of EW(H$\alpha$). ii) Non-interacting galaxies show redder colours and lower EW(H$\alpha$) values; most of them harbour an HII or Sy1 type nucleus.
iii) Our regions seem to show values that match well, in the EW(H$\alpha$)-colour diagrams, (Figs \ref{EW-blue} and \ref{EW-red}) with those shown by the stars expected to be present in HII regions or objects alike, taking notice that we reach lower H$\alpha$ equivalent width values. However, in the colour-colour diagram (Fig \ref{colour-colour}) our regions seem to segregate from the comparison sample, showing redder colours.
If this figure is considered together with \ref{EW-red} even though only

Figure 7 must be considered together with figure 11. Even though only markarian objects are in common, it can be checked that while in  the EW line vs red colours data overlap, this effect disappears when a pure (U-B) colour is introduced. The cause of this split effect, could be a second redder poorly or non ionizing population of stars that is  present in some individual HII regions.

We can attribute this latter behaviour to four main reasons:

\begin{itemize}
\item[-]{An age effect, the EW(H$\alpha$) diminishes as the burst gets older and it also appears redder.}

\item[-]{A metallicity effect. In the colour-magnitude (I vs R-I) of Fig \ref{i-ri}, only NGC 3310 is segregated from the other galaxies. NGC 3310 is known to be a low metallicity object, so the others must have solar o oversolar metallicities. (\citealt{pas:93}; \citealt{dia:07})}

\item[-]{There could be a mismatch in the estimation of extinction, although computed values agree within errors with those available in the literature.}

\item[-]{We may be detecting a composite stellar population. Under this scenario, a very young population would be responsible for most of the EW(H$\alpha$), while a population older than 10 Myr would be the source of the red continuum and responsible for the detection of EW(H$\alpha$) values smaller than those expected for a single burst of ionizing stars.}
\end{itemize}

Besides the above, some remarks on individual objects can be made.
HII regions in NGC 4321 show low U emission but average H$\alpha$ luminosities due maybe to an additional ionization source rather than only young stars.
Even more interesting are the regions with no peculiar behaviour in luminosity or colours but with unexpected  low H$\alpha$ EW, as for instance those located in NGC 4314. ¿Are we observing overlapping stellar populations of different ages? To disentangle between effects than can be present simultaneously in a galaxy, and to achieve a better understanding of the nature and evolution of every individual HII region, a closer insight to each of them, in the light of theoretical evolutionary models, is needed. This will be presented in a forthcoming paper.

\section{Summary and conclusions}
From high quality continuum subtracted $H\alpha$ and $H\beta$, and U, B, R, I broad band images of a sample of 20 galaxies with different degrees of nuclear activity and circumnuclear star formation, we have catalogued a total of 336 HII regions. For all of them  $H\alpha$ and $H\beta$ luminosities and equivalent widths have been measured, together with R-I and  U-B colours and extinction values.

Trying to settle whether a ring can be considered circumnuclear or not, we have defined a ``limiting radius'' as a function of the galaxy absolute magnitude. Given the galaxy integrated absolute magnitude (B band) this radius can be simply computed, so HII regions beyond this radius will not be misclassified as circumnuclear.

Using $\frac{H\alpha}{H\beta}$ theoretical values, together with observed narrow H$\alpha$ and H$\beta$ images, has allowed us to correct colours from extinction and disentangle age and metallicity effects. The visual extinction values found (A$_V$) range from 0 to 4.75 magnitudes with an average value of 1.85, in good agreement with other published values. We have found no correlation between extinction and galactocentric distance or region size. We have found loose correlations between extinction and luminosities or colours, with redder compact sources showing higher extinction values.

Circumnuclear HII regions are known to be very bright, the H$\alpha$ luminosities range we have found, varies from  $1.3 \times 10^{38}$ to $4\times 10^{41} erg s^{-1}$, with most of the regions showing values of log[L(H$\alpha)] \geq$ 39, therefore we are mostly dealing with giant HII regions. No clear correlation seems to exist between H$\alpha$ luminosity and other galactic features as morphology or nuclear activity of the host galaxy.

Studying the behaviour of the H$\alpha$ luminosity function and the dependence of luminosity on the radius of each region, we infer that, at least some of the biggest and less luminous regions, could be matter bounded.

Only 10\% of the regions show EW(H$\alpha$) values larger that 250 \AA. Their host galaxies are either Sy2 or LINERs. Values of EW(H$\alpha$) larger than 450 \AA, are only found in interacting galaxies. 
Broad band colour distribution has been found to be bimodal, with the bluest regions being located in galaxies showing evindence for interactions and LINER or Sy2 activity in their nuclei. These bluer regions are, on average, also stronger H$\alpha$ emitters and show high EW(H$\alpha$) values.

In summary, HII regions hosted by interacting galaxies with LINER or Sy2 activity, seem to be bluer and show higher values of EW(H$\alpha$), on average, than HII regions located in the other types of galaxy. This fact suggests younger ages for the former regions if the metallicity of the galaxies is similar. 

On the other hand, in colour-magnitude, colour-colour and EW(H$\alpha$)-colour diagrams our HII regions appear to segregate from a comparison sample of simple stellar populations known to be present in young HII regions.
This fact may be attributed to age, metallicity or mismatching extinction corrected effects. 
It may also be due to the presence of a second young but non-ionizing population of stars, that would make the ionizing burst appear redder. The continuum coming from this second older burst would also be responsible for low values of the  EW(H$\alpha$).

To study whether or not this second non-ionizing burst is present, and in which proportion for each cluster, a closer look into the nature of HII regions in the light of theoretical evolutionary models is proposed in a forthcoming paper.\\

\medskip 

\textbf{Appendix}\\

\textbf{Notes on individual objects}\\

\textbf{NGC 1068}\\
Being the strongest nearby Seyfert 2 galaxy, NGC 1068 has been extensively observed over the whole electromagnetic spectrum. \cite{ant:85} detected broad permitted lines in its polarized spectrum, becoming clear that a hidden broad line region should exist. Since then, NGC 1068 has become a representative example of the so called AGN Unified Model.
From the inside out, several galactic structures have been observed: (a) 
A  1 pc compact hot dust source in the nucleus measured using near-infrared speckel imaging and integral field spectroscopy (\citealt{tha:97}); (b)  A central circumnuclear disk with an aproximate diameter of 300 pc observed in CO and HCN molecules (\citealt{sci:00}; \citealt{tac:94}); (c) At a radius of 1-1.5 Kpc a prominent starburst ring has been observed at molecular, IR and visible wavelengths   (\citealt{sci:00}; \citealt{tac:94}; \citealt{sco:88}). A strong CaII triplet has also been observed and attributed to red supergiant stars (\citealt{ter:90}; \citealt{dre:84}); (d) 
A 2.3 Kpc stellar bar is also clearly revealed by Near-IR observations (\citealt{sco:88}); (e)
A jet extensively observed from centimeter wavelengths to IR, can be seen up to distances of several Kpc from the nucleus.
	The fact that all these features occur simultaneously in this object makes of it an excellent laboratory for the study of possible interrelations between them. \\

\textbf{NGC 2782}\\
NGC 2782 (Arp 215) is an isolated early-type spiral galaxy [SAB(rs)a], whose optical and HI signatures
(\citealt{kri:05}) indicate a possible merging episode between galaxies about 200 Myr ago (\citealt{zan:06}).
The central region of NGC 2782 hosts a nuclear starburst and radio and X-ray observations seem to point to the existence of an optically hidden AGN.\\
 	It is well established that large-scale bars transport gas inward very efficiently, driving powerful starbursts in timescales of few times $10^{8}$ yr but, given that AGN accretion occurs in much shorter timescales in the life of a galaxy, it is less likely to observe a galaxy in its AGN intermittent period than to observe gas infall.  NGC 2782 could be one of the rare cases of a galaxy with this nuclear accretion \lq\lq{switched on}\rq\rq
(\citealt{sai:94}).
	Three-dimensional optical spectroscopy
(\citealt{yos:99}; \citealt{gar:02})
shows evidence for high-speed ionized gas outflows with bipolar structure, and there is also evidence of high-excitation extranuclear emission lines due to shock heating (\citealt{boe:92}).\\
\cite{gar:02} also detected in colour maps the region we have labeled as number 2, at the East of the nucleus and extending about 5$^{\prime\prime}$. They consider it an especially reddened region. We found for this region Av=2.6, the maximum value for all HII regions in NGC 2782 while the mean Av value for HII regions in this galaxy is 1.08.\\

\textbf{NGC 2903}\\
NGC 2903 is a very luminous SAB(rs)bc galaxy that possesses a complex structure in its inner region where several hot spots are clearly resolved, together with a symmetric strong bar.
These hot spots where detected for the first time by S\'ersic \& Pastoriza (1967). Some of them present emission lines and colours typical of hot star populations but others do not show optical emission lines, so they must be formed by older stars. \\
	Alonso-Herrero, Ryder and Knapen (2001) reported lack of spacial correspondence between the bright HII regions seen in Pa$\alpha$ images and the position of stellar clusters in the H-band. It appears that the bright NIR sources, although close to the HII regions are not coincident with the maximum H$\alpha$ emission. They interpret this fact in the sense that bright IR stellar clusters are the result of the evolution of HII regions.
By combining GEMINI data and a grid of photo-ionization models \cite{dor:08}, conclude that the contamination of the continua of CNSFRs by underlying contribution from both old bulge stars and stars formed in the ring in previous episodes of star formation (10-20 Myrs) yield the detected low equivalent widths that we also observe.
H\"agele et al. (2009) derived gas and star velocity dispersions for four CNSFR and the nucleus of NGC 2903 using high resolution spectroscopy in the blue and far red, finding results compatible with composite stellar populations for the knots.\\
	A multiwavelength study of star formation in the bar and inner parts of NGC 2903 by Popping, P\'erez \&  Zurita (2010), shows that the CO(J=1-0) and the 2-3 $\mu$m emission trace each other in a barred shape. H$\alpha$ has an s-shape distribution and the UV is patchy, symmetric with respect to the galactic centre, and does  not resemble the bar. The ages estimated range from less that 10 Myr for the H$\alpha$ emitting knots to 150 to 320 Myr for regions identified by their UV emission.\\
	Region labeled as 2 in the present work, has been identified as 1 by Alonso-Herrero, Ryder \& Knapen (2001), finding good agreement in Av (1.7 our value vs 1.8 theirs) and a slightly higher $H\alpha$ luminosity.
 H\"agele et al. (2009)  identified regions 2 and 1 as R4 and R7.  The values they find (see table 6 of their work and references therein) for extinction match with ours.\\
	Popping, P\'erez \&  Zurita (2010) identified the following regions:\\
Present work region label/Popping, P\'erez \&  Zurita (2010) region label: 4/5; 5-6/6; 7/55; 8/13; 9/3; 11/51; 12/20; 13/21; 14/4; 15/15; 16/14; 17/2; 18/65; 21/66.                  
The values that we find for H$\alpha$ luminosities are slightly higher. Regarding the EW of this line, the values we measure are about six times smaller, and good agreement is found for extinction values in regions labeled as 9, 11, 12, 13 and 17 in our work. For the rest of the regions our value is higher except in region 7.\\

\textbf{NGC 3310}\\
NGC 3310 (Arp217, UGC 5786) is a SAB(r)bc starburst galaxy with an inclination of the galactic disc of $ i \sim 40^{o}$ as computed by S\'anchez-Portal et al. (2000), at a distance of 12.5 Mpc. 
This galaxy exhibits a ring of  giant HII regions at an average distance of 700 pc from its nucleus (\citealt{elm:02}; \citealt{van:76}).
These circumnuclear regions, in contrast to what is generally found in this type of objects (\citealt{dia:07}), show low metallicity values.\\
	\cite{pas:93} and H\"agele et al. (2010) presented high-resolution far-red spectra and stellar velocity dispersion measurements along the line of sight for eight knots located in the circumnuclear ring of this galaxy. H$\alpha$ and H$\beta$ equivalent widths were also computed together with broad-band V, R and I photometry by D\'\i az et al (2000), finding good agreement with our data. \\

\textbf{NGC 3949}\\
NGC 3949 is a SA(s) bc galaxy with no detected nuclear activity. It shows a very luminous nucleus surrounded  by some HII regions barely resolved\\.

\textbf{NGC 3982}\\
NGC 3982 has been classified as a Sy2 galaxy due to its emission line espectrum (\citealt{hof:97}), although its continuum ultraviolet light is dominated by the stellar emission of a non Seyfert galaxy (\citealt{kin:93}) showing evidence for stellar absorption lines. In 1997 Gonz\'alez-Delgado et al. detected [OIII] emission in the nucleus itself and in the circumnuclear region. Very interesting is also the fact that Ho et al. (1997), claim to have observed a broad component in H$\alpha$.\\

\textbf{NGC 4314}\\
NGC 4314 contains one of the biggest nuclear rings of star formation that can be observed in barred galaxies and early type spirals. In spite of its symmetrical aspect it was classified as peculiar by Sandage (1961) due to the presence of dust lines and the peculiar structure of its central bar. \\
	UBV observations carried out by \cite{lin:73}  show bluer colours towards the centre of the galaxy, probing that, at least in this inner region, recent star formation has been very intense. Observations of [H$\alpha$ + NII] (Burbidge \& Burbidge 1962; Wakamatsu \& Nishida (1980) confirmed this high rate of star formation.
\cite {ben:02} after a series of observations of NGC 4314 that include CO radio observations,  U, B, V, I, and H$\alpha$ HST photometry of 80 star clusters in the circumnuclear ring, concluded that the blue colours and strong H$\alpha$ emission of some of these clusters may suggest very young ages, less than 5 Myr in some cases. The EW of the H$\alpha$ line that we observe is also consistent with these young ages. Just exterior to this ring of extremely young stars, colours are consistent with stellar ages ranging from 100 to 300 Myr suggesting some kind of age evolution in the ring and its surroundings. Although our observations have lower resolution, the (U-B) colour can be compared in several cases finding good agreement. Our values range from -0.29 to -0.96 versus 0 to -1.01 in their work. As an individual example our region labelled 1 (U-B = -0.75), includes regions 51 (U-B = -0.75), 53 (U-B = -0.72), 54 (U-B = -0.84) and 57 (U-B = -0.87) in their work.\\
	\cite {ben:02}) found average values of Av for HII regions in this galaxy smaller than ours. This fact does not affect any conclusions based on average luminosities and colours.\\

\textbf{NGC 4321}\\
Also known as M100, this galaxy  is the biggest of the Virgo cluster. Its nuclear region has been well studied in the optical (\citealt{pie:86}) and CO emission (\citealt{ran:95}; \citealt{sak:95}). These studies have unveiled an intricate morphology associated with the intense star formation that is taking place in the inner kiloparsec. A ring of bright HII regions is surrounding this central part with a mean radius of about 1 kpc. The H$\alpha$ line emission is twice more intense than that of the nucleus itself (\citealt{ars:88}). Ryder \& Knapen (1999) measured colours and magnitudes in the I, J, H and K bands for 41 circumnuclear HII regions of this galaxy, concluding that data are consistent with the emission by a population of stars with ages ranging from 15 to 25 million years, dominated by supergiant M0 stars.\\

\textbf{NGC 5033}\\
NGC 5033 is a nearby spiral galaxy that contains a Seyfert nucleus (\citealt{hof:97}) and has been reported to host at least three supernovae events in  the last 60 years (SN 1950C, SN 1985 L and SN 2001 gd). Thean (1997), using Very Large Array (VLA) images measured a total HI mass of $7.0 ~ 10^9 M_{\odot}$, and P\'{e}rez-Torres \& Alberdi (2007) presented continuum VLA observations at 4.9 and 8.4 GHz of the inner 40 Kpc of the galaxy. They came to the conclusion that the radio and FIR emission of NGC 5033 seems to be dominated by a 10 Myr old starburst.\\
	Terashima (1999) reported the detection of an X-Ray source at the nucleus of the galaxy. \cite{kot:03}; using aperture synthesis observations of CO(J=1-0) emission and near infrared broad-band photometry, suggested the presence of a nuclear bar. Integral field spectroscopy carried out by \cite {med:05}, support the conclusion of the presence of non symmetric departures in the gravitational potential which could be fuelling the active nucleus.\\

\textbf{NGC 5194}\\
NGC 5194 (the Whirlpool Galaxy, M51) together with its companion NGC 5195, forms one of the best studied pair of galaxies at all wavelengths. It is one of the nearest grand-design (SAbc) spiral that shows a prominent two arm configuration. A panchromatic study by Calzetti et al (2010), using GALEX, Spitzer, HST and ground-based observations, yielded a SFR surface density of $ 0.015  M_\odot$ yr$^{-1} $Kpc$^{-2} $ and a decreasing dust extinction as a function of galactocentric distance, with maximum central values of Av = 3.5 mag.
D\'{i}az et al. (1991) and, later, Moustakas et al. (2010)  classified it as metal rich from spectroscopic measurements.\\

\textbf{NGC 5248}\\
NGC 5248 is a nearby grand-design spiral galaxy with a spiral structure that extends from 0.1 to 10 Kpc from its nucleus.
Jogee et al (2002) have shown from deep R-band images and dynamical and photometric analyses that the spiral structure is being driven by a moderately strong stellar bar. The inner kiloparsec of the bar hosts a well known ring of ``hot spots'' that has been resolved into bright HII regions and young star clusters (\citealt{mao:01}).
A second nuclear ring with a radius of 1$^{\prime\prime}$.25 and a nuclear dust spiral has also been observed by \cite{lai:01}.
This complex inner structure has been confirmed in a more recent study made by Yuan and Yang (2006).\\

\textbf{NGC 5929/5930}\\
NGC 5929 and NGC 5930 form a pair of nearby interacting galaxies (\citealt{ken:87}). Space and ground based observations have confirmed the presence of ionized gas in both galaxies. For NGC 5929 a Seyfert nucleus has been observed, together with a bi-polar radio jet (\citealt{ros:10}). It has an absorption-corrected hard X-ray luminosity which is low compared to average local Seyferts (\citealt{car:07}).
On the other hand, the ionized gas detected in NGC 5930 seems to have its origin in a ring of circumnuclear star forming regions (\citealt{bow:95}).\\

\textbf{NGC 5953/5954}\\
NGC 5953 and NGC 5954 are a pair of galaxies separated by a projected distance of 6 Kpc that shows clear signs of interaction. Both galaxies show emission associated with a circumnuclear burst of star formation. NGC 5953 has been classified as a Seyfert galaxy hosting a compact radio core and a ``jet'' first referred to by \cite{gon:96}; and NGC 5954 shows LINER features.
\cite{her:03} using BVRIH$\alpha$JK and [NII] scanning Fabry-Perot observations, derived surface brightness profiles for both galaxies and proposed a geometrical scenario for the encounter.
Radio to  X-ray Chandra observations published by \cite{ros:08} suggest the possible presence of a binary black hole with a mass over 50 $M_{\odot}$, associated with a young star cluster.
Casasola et al. (2010), mapped the molecular gas of NGC 5953 using high resolution observations of the CO line taken with the IRAM interferometer, together with IR and HST archive images. They suggest that there is an apparent counter-rotation between gas and stars inside and outside the ring, with the ring constituting the border line between both dynamical components. The AGN in NGC 5953 is apparently not been actively fuelled in the current epoch.\\

\textbf{NGC 6574}\\
NGC 6574 is a nearby Seyfert2 galaxy with not many references in the literature. \cite{lin:08} analyzed interferometric data obtained with IRAM in the CO(1-0) and CO(2-1) lines, finding that the molecular gas in this galaxy is distributed in four components: nucleus, bar, spiral arms (pseudo-ring), and extended underlying disk component. \cite{sak:99} found a prominent CO peak of 1 Kpc in diameter at the centre of the galaxy.
\cite{kot:00} and \cite{lai:01} present high spatial resolution near infrared observations, finding good agreement between radio and infrared morphologies for the circumnuclear star forming ring and proposing an instantaneous burst of star formation occurring 6-7 Myr ago to explain the data. These ages are also derived from H$\alpha$ observations (Gonz\'{a}lez-Delgado et al. 1997).\\

\textbf{NGC 6951}\\
NGC 6951 was first classified as a LINER type nucleus but recent works  \citep{per:00} suggest that it shows an intermediate activity between LINER and Seyfert.
IRAM CO(1-0) and CO(2-1) observations of the galaxy in the NUGA project (\citealt{van:11}) show that the CO circumnuclear emission arises from two different components: the gas from the bar and a high velocity component likely connected to the nuclear stellar oval. This last component may represent the last link in the chain of gas inflow towards the Active Galactic Nucleus of NGC 6951.
These streaming motions along the nuclear spiral arms towards the nucleus, have also been concluded by \cite{sto:07} with observations of the ionized gas in the inner region of the galaxy and by \cite{roz:02}.\\

\textbf{NGC 7177}\\
NGC 7177 is a barred galaxy with LINER activity in its nucleus. \cite{hod:82} observed its circumnuclar ring of star forming regions and it is also included in some CO and ionized gas catalogs (\citealt{ver:85}; \citealt{pog:89})\\

\textbf{NGC 7469}\\
NGC 7469 is one of the brightest blue galaxies first listed by Seyfert (1943). Its nucleus is surrounded by a star forming ring well studied by \cite{dis:07} using HST UV through NIR images together with K-band ground-based long slit spectroscopy. They also collected existing IR and radio maps from the literature. They found that two different populations of massive clusters coexisted in the inner regions with ages ranging from 1-3 Myr for the youngest to 9-20 Myr for the oldest. A compound stellar population of these two ages is also consistent with the observed EW and colours that we have computed in the present work.\\
	NGC 7469 has also a large stellar bar detected in the NIR (\citealt{kna:00}) and a nuclear bar with a size similar to the ring (\citealt{dav:04}).
The supernova event SN2000ft has recently been detected on the ring and its optical counterpart has been identified (\citealt{col:01}; \citealt{alb:06}; \citealt{col:07})\\

\textbf{NGC 7714}\\
Among nearby interacting non active galaxies, the pair NGC 7714/15 (Arp 284) is one of the best studied. Three circumnuclear regions were observed by \cite{gon:95}) with long-slit spectroscopy finding oxygen abundances of about half of the solar value and WN features that constitute evidence of very young ages. A deeper study of the stellar populations of these regions can be found in Garc\'\i a-Vargas et al. (1997) and the young stellar component of the central 300 pc region of the galaxy has been studied by Gonz\'alez-Delgado et al. (1999) and \cite{lan:01} by means of ultraviolet spectroscopy with the HST/GHRS and ground-based optical data finding an average age of about 4.5 Myr, with little evidence for an age spread. More recently NGC 7714 has been observed in X-Rays by \cite{smi:05} finding that the nuclear HII regions are the source of a significant fraction of the emission detected. Again, stellar population ages below 4 Myr are needed to explain the data. These extremely young stars seem to be the source of the high values of equivalent widths of the H$\alpha$ line that we also observe.
Spitzer observations of NGC 7714 carried out by \cite{bra:04} confirmed the HII-line spectrum of the nucleus of this galaxy, with no evidence of an obscured active galactic nucleus. These authors have also detected strong polycyclic aromatic hydrocarbon emission features.\\

\textbf{ACKNOWLEDGEMENTS}\\

We wish to thank an anonymous referee for suggestions that greatly improved the clarity of the paper.\\

Financial support for this work has been provided by the Spanish \textit{Ministerios de Educaci\'{o}n y Ciencia and Ciencia e Innovación}  under grants AYA2007-67965-C03-03 and
AYA2010-21887-C04-03,  and the \textit{Junta de Comunidades de Castilla-La Mancha}. ET and RT are also grateful to the Mexican Research Council (CONACYT) for support under grants CB2005-01-49847F and CB2008-01-103365.

This research has made use of the NASA/IPAC Extragalactic Database (NED) which is operated by the Jet Propulsion Laboratory, California Institute of Technology, under contract with the National Aeronautics and Space Administration.

The 1.5 m-telescope of Calar Alto Observatory is operated under the control of the Spanish National Astronomical Observatory (OAN).  We thank the Spanish allocation committee (CAT) for awarding observing time.

\end{document}